\documentclass[fleqn,usenatbib]{mnras}
\usepackage{newtxtext,newtxmath}
\usepackage{comment}
\usepackage{multirow}
\usepackage{mathtools}
\usepackage[T1]{fontenc}
\usepackage{ae,aecompl}
\usepackage{graphicx}	
\usepackage{cleveref}	
\usepackage[ruled,vlined]{algorithm2e}
\usepackage{algpseudocode}
\usepackage{enumitem} 
\usepackage{xfrac} 

\setlist[itemize]{leftmargin=*}
\newcommand{\bigO}[1]{\mathcal{O}(#1)}

\newcommand{\norm}[1]{\|#1\|}

\newcommand{\ud}{\mathrm{d}}

\newcommand{\vect}[1]{\boldsymbol{#1}}

\DeclarePairedDelimiter\floor{\lfloor}{\rfloor}
\newcommand{\derfrac}[2]{\frac{\ud #1}{\ud #2}}
\newcommand{\pderfrac}[2]{\frac{\partial #1}{\partial #2}}

\newcommand{\frost}{\texttt{FROST}}
\newcommand{\op}[1]{{\vect{\mathrm{#1}}}}

\title[A hierarchical forward symplectic integrator]{{
\Huge{\texttt{FROST}}}: a momentum-conserving CUDA implementation of a hierarchical fourth-order forward symplectic integrator}
\author[A. Rantala et al.]{Antti Rantala$^{1}$\thanks{E-mail: anttiran@mpa-garching.mpg.de}, Thorsten Naab$^{1}$, Volker Springel$^{1}$\\
$^{1}$Max-Planck-Institut f\"ur Astrophysik, Karl-Schwarzschild-Str. 1, 
D-85748, Garching, Germany\\
}
\date{Accepted XXX. Received YYY; in original form ZZZ}
\pubyear{2020}

\begin{document}
\label{firstpage}
\pagerange{\pageref{firstpage}--\pageref{lastpage}}
\maketitle

\begin{abstract}
We present a novel hierarchical formulation of the fourth-order forward symplectic integrator and its numerical implementation in the GPU-accelerated direct-summation N-body code \frost{}. The new integrator is especially suitable for simulations with a large dynamical range due to its hierarchical nature. The strictly positive integrator sub-steps in a fourth-order symplectic integrator are made possible by computing an additional gradient term in addition to the Newtonian accelerations. All force calculations and kick operations are  synchronous so the integration algorithm is manifestly momentum-conserving. We also employ a time-step symmetrisation procedure to approximately restore the time-reversibility with adaptive individual time-steps. We demonstrate in a series of binary, few-body and million-body simulations that \frost{} conserves energy to a level of $|\Delta E / E| \sim 10^{-10}$ while errors in linear and angular momentum are practically negligible. For typical star cluster simulations, we find that \frost{} scales well up to $N_\mathrm{GPU}^\mathrm{max}\sim 4\times N/10^5$ GPUs, making direct summation N-body simulations beyond $N=10^6$ particles possible on systems with several hundred and more GPUs. Due to the nature of hierarchical integration the inclusion of a Kepler solver or a regularised integrator with post-Newtonian corrections for close encounters and binaries in the code is straightforward. 
\end{abstract}



\section{Introduction}\label{section: 1}

Gravitational direct N-body simulations of collisional star clusters have recently reached the million-body era (e.g. \citealt{Wang2016}). The standard time integration procedure in such simulations during the past few decades has been the fourth-order Hermite scheme \citep{Aarseth1999}, while even higher-order Hermite integrators exist \citep{Nitadori2008}. This fourth-order scheme is a predictor-corrector integrator based on third-order force polynomials constructed from particle accelerations and their time derivatives \citep{Makino1992, Hut1995, Aarseth2003}.

The Hermite integrator is typically accompanied by a neighbour scheme separating the rapidly evolving short-range forces and the slowly changing long-range forces \citep{Ahmad1973} as well as a block time-step scheme sorting the particles into a factor of two hierarchy according to their individual time-steps \citep{Mclachlan1995,Hernquist1989,Makino1991}. Hard binaries and close particle encounters are often \citep{Aarseth2003, Mikkola2008a}, but not always \citep{Konstantinidis2010, Hubber2018}, integrated with specialised regularisation techniques \citep{Kustaanheimo1965, Mikkola1993, Preto1999, Mikkola1999, Mikkola2008}.

In addition to algorithmic improvements, particle numbers in direct summation simulations with the Hermite integrator have been increasing due to the development of special-purpose hardware like GRAPE \citep{Ito1990,Makino2008} and the efficient use of general purpose many core accelerators (graphics processing units, GPUs) in astrophysical high-performance computing (\citealt{Gaburov2009,Nitadori2012,Wang2015}).

While Hermite codes have become the standard for collisional N-body simulations, alternative numerical algorithms for directly integrating the gravitational N-body problem have been explored \citep{Dehnen2011}. Here, symplectic integrators \citep{Yoshida1990,Yoshida1993} are of particular interest. By employing the geometrical properties of Hamiltonian mechanics \citep{Hairer2006} symplectic integrators preserve the Poincar\'{e} integral invariants i.e. the phase-space of the dynamical system. They also exactly conserve the so-called surrogate Hamiltonian $\tilde{H}$ close to the original Hamiltonian $H$ as
\begin{equation}\label{eq: surrogate_hamiltonian_intro}
    H = \tilde{H} + H_\mathrm{err}
\end{equation}
in which $H_\mathrm{err}$ is the so-called error Hamiltonian characterising the typically small difference of the surrogate Hamiltonian and the original Hamiltonian. The conservation of the surrogate Hamiltonian very often yields good energy conservation, especially for long-term integrations. Despite their high accuracy per integration step the widely-used Hermite integrators are not symplectic in nature and may be susceptible to long-term secular error growth \citep{Binney2008, Dehnen2011}.

Symplectic integrators are constructed using Hamiltonian splitting. In general a Hamiltonian $H$ is separable if it can be expressed as a sum of two parts $H = H_\mathrm{A} + H_\mathrm{B}$ in which $H_\mathrm{A}$ depends only on the canonical coordinates and $H_\mathrm{B}$ on the corresponding momenta of the particles of the dynamical system. For separable Hamiltonians first and second-order symplectic integrators, the Euler integrator and the leapfrog integrator, can be constructed. Moreover, if the second-order leapfrog exists the seminal method of \cite{Yoshida1990} allows for the construction of higher-order symplectic integrators for any even order.

A common procedure in Hamiltonian mechanics is the splitting of the Hamiltonian $H$ into kinetic $T$ and potential $U$ parts as $H=T+U$. The kinetic term $T$ generates the drift operator $e^{\epsilon \op{T}}$ which propagates the coordinates of the dynamical system over a time interval $\epsilon$. The kick operator $e^{\epsilon \op{U}}$ generated by the potential term $U$ updates the momenta i.e. velocities if the masses of the elements of the dynamical system are constant.

The separation of the Hamiltonian into kinetic and potential parts is not the only option when constructing symplectic second-order N-body integrators. In N-body systems with a dominant gravitating body the Wisdom-Holman splitting separates the Hamiltonian into $N-1$ Keplerian two-body Hamiltonians $H_\mathrm{i}^\mathrm{Kepler}$ between the dominant body and other particles and $(N-1)(N-2)/2$ perturbative interaction Hamiltonians $U_\mathrm{ij}^\mathrm{int}$ between the non-dominant bodies \citep{Wisdom1991,Murray2000,Hernandez2015,Rein2015,Rein2019}. Specialised numerical techniques have been developed for the perturbed Keplerian Hamiltonians (e.g. \citealt{Danby1992,Hernandez2015,Wisdom2015,Dehnen2017a,Hernandez2020,Rein2020}). These integrators are widely used in the context of gigayear-long simulations of Solar system bodies.

Yet another class of symplectic integrators can be derived using hierarchical Hamiltonian splitting (hereafter HHS). The HHS provides an attractive alternative to the widely-used block time-step scheme for simulations with individual particle time-steps \citep{Saha1994, Pelupessy2012, Janes2014}. Starting from a pivot time-step $\tau_\mathrm{pivot}$ the Hamiltonian is adaptively divided into Hamiltonians $H_\mathrm{S}$ and $H_\mathrm{F}$ of slow ($\tau_\mathrm{i} \geq \tau_\mathrm{pivot}$) and fast particles ($\tau_\mathrm{i} < \tau_\mathrm{pivot}$) according to the individual time-steps $\tau_\mathrm{i}$ of the particles. The process is repeated recursively on $H_\mathrm{F}$ with increasingly smaller pivot time-steps until no particles remain in the set of fast particles. On a single hierarchy level the Hamiltonian splitting then is
\begin{equation}\label{eq: intro-fastslow}
    H = H_\mathrm{S} + H_\mathrm{F} + H_\mathrm{SF} = H_\mathrm{S} + H_\mathrm{F} + U_\mathrm{SF},
\end{equation}
in which $U_\mathrm{SF}$ is the interaction Hamiltonian between the sets of slow and fast particles. Thus at the end of the HHS procedure only a collection of slow Hamiltonians and interaction Hamiltonians remains. The number of these Hamiltonians depends on the time-step distribution $\{\tau_\mathrm{i}\}$ of the particles. The HHS does not constrain how the particle time-steps should be chosen so the time-step assignment is a separate choice to be made.

The interaction Hamiltonians $U_\mathrm{SF}$ between the hierarchy levels ensure that inter-level force calculations and corresponding kick operations are always pair-wise, which is not true for the block time-step scheme. Thus integrators derived using the HHS are manifestly momentum-conserving. The interaction Hamiltonian $U_\mathrm{SF}$ can be placed on the same hierarchy level as the corresponding slow Hamiltonian $H_\mathrm{S}$ rendering the dynamics of the fast particles generated by $H_\mathrm{F}$ independent of the slower hierarchy levels. This remarkable decoupling of rapidly evolving dynamical sub-systems enables efficient integration of systems with an extreme dynamical range (e.g. \citealt{Pelupessy2012,Zhu2020,Springel2020,Mukherjee2020}).

A common property for all symplectic integrators beyond the second order derived by using the method of \cite{Yoshida1990} is the unavoidable occurrence of negative integration sub-steps. While negative time-steps are not a problem for Newtonian gravitational dynamics due to its time-reversibility, they cause problems for important time-irreversible dynamical processes such as gravitational-wave emission and tidal dissipation. In addition, negative time-steps make the attractive higher-order hierarchical integration methods prohibitively inefficient \citep{Pelupessy2012}.

A rather original and surprisingly rarely used solution to avoid negative integration steps in a fourth-order symplectic integrator is to move appropriate terms from the error Hamiltonian $H_\mathrm{err}$ into the surrogate Hamiltonian $\tilde{H}$ in Eq. \eqref{eq: surrogate_hamiltonian_intro}. This process results in a family of forward symplectic integrators (hereafter FSI) which contain only positive integration sub-steps at the cost of evaluating the force gradient in addition to the Newtonian force term \citep{Chin1997,Chin2005}. Even though fourth-order forward integrators have been proven to be extremely efficient and accurate in few-body gravitational dynamics \citep{Chin2007} they have not been widely adopted by the astrophysical community \citep{Dehnen2011}. To the best of the authors' knowledge the only N-body implementation of the forward integrator is the \texttt{TRITON} code \citep{Dehnen2017a} which also uses a specialised Kepler solver.

In this article we describe a novel integration method HHS-FSI which combines a hierarchical integration scheme with the fourth-order forward symplectic integrator. First, our new symplectic integrator is derived by using the HHS technique enabling the efficient integration of systems with extremely large dynamical ranges. Next, the inter-particle force calculations in the code are always pair-wise making the algorithm manifestly momentum-conserving. We employ a forward symplectic integrator and a novel fourth-order Hamiltonian split on Eq. \eqref{eq: intro-fastslow} rendering the entire integration algorithm fourth-order accurate. Finally, in contrast to most available higher-order symplectic integrators our technique contains no negative integration steps. 

The HHS-FSI integrator is implemented in the novel N-body code \frost{}. The code is written in MPI-parallelised CUDA C language to enable the use of hardware-accelerated computation nodes in modern CPU-GPU computing clusters and supercomputers. Pseudocode instructions on implementing a version of the HHS-FSI integrator are provided as a part of this study.

This work is organised as follows. In Section \ref{section: 2} we review the construction and implementation of the standard fourth-order forward symplectic integrator. In Section \ref{section: 3} we describe the hierarchical Hamiltonian splitting technique and present the novel hierarchical fourth-order forward symplectic integrator. Time-stepping used with the integrator is presented in Section \ref{section: 4} and the numerical implementation of the integrator in Section \ref{section: cuda}. The order and numerical accuracy of the integrator as well as running speed and scaling of the \frost{} code are validated by various numerical experiments in Section \ref{section: 6}. Appendixes A1 and A2 provide details of the initial conditions for the simulations in this Section. Finally, we summarise our main results and conclude in Section \ref{section: 7}. 


\section{Forward symplectic integrators}\label{section: 2}

\subsection{Symplectic integrators}

In this Section we review symplectic integration methods (e.g. \citealt{Yoshida1990,Yoshida1993}) and their derivation up to fourth order with the overall goal of presenting the fourth-order forward symplectic integrator \citep{Chin1997,Chin2005,Chin2007,Dehnen2017a}. The integrator is not very well-known and has not been widely used in N-body studies despite its suitability for accurate orbital integration \citep{Dehnen2011}. We largely follow the notation of \cite{Dehnen2017a} in this article.

The Hamiltonian equations of motion of a dynamical system can be written as a single equation using the compact notation as
\begin{equation}
\derfrac{\vect{w}}{t} = \op{H} \vect{w} = \Big\{ \vect{w}, H \Big\}_\mathrm{P}
\end{equation}
in which $\vect{w} = \{ \{\vect{x}_\mathrm{i} \}, \{ \vect{p}_\mathrm{i} \} \}$ is the phase space state of the dynamical system and $\{,\}_\mathrm{P}$ are the Poisson brackets. The operator $\op{H}$ is the so-called Lie operator of the Hamiltonian $H$ \citep{Dragt1976}. The Hamiltonian equation of motion has a formal solution over a time interval $\epsilon$ as
\begin{equation}\label{eq: H-formal-solution}
\vect{w}(t+\epsilon) = e^{\epsilon \op{H}}\vect{w}(t)
\end{equation}
in which $e^{\epsilon \op{H}}\vect{w}(t)$ is the time evolution operator generated by $H$ \citep{Goldstein1980}. Symplectic integrators are derived (e.g. \citealt{Dehnen2017a}) from proper continuous canonical transformations of Eq. \eqref{eq: H-formal-solution}. Consequently, symplectic integrators preserve the phase space i.e. Poincar\'{e} integral invariants of the dynamical system \citep{Hairer2006}. 

A time evolution operator $e^{\epsilon \op{H}}$ generated by a separable Hamiltonian $H=T+U$ can be decomposed into an operator product
\begin{equation}\label{eq: Hamiltonian_product_1}
    e^{\epsilon \op{H}} = e^{\epsilon (\op{T}+\op{U})} = \prod_\mathrm{i=1}^\mathrm{N} e^{\epsilon t_\mathrm{i} \op{T}} e^{\epsilon u_\mathrm{i} \op{U}}
\end{equation}
if the individual drift and kick operators
\begin{equation}
\begin{split}
&e^{\epsilon \op{T}}\{\vect{x}_\mathrm{i}(t)\} = \{\vect{x}_\mathrm{i}(t+\epsilon)\} \hspace{1cm}\text{drift}\\
&e^{\epsilon \op{U}}\{\vect{p}_\mathrm{i}(t)\} = \{\vect{p}_\mathrm{i}(t+\epsilon)\}\hspace{0.93cm}\text{kick}
\end{split}
\end{equation}
can be exactly computed. The set of coefficients $\{t_\mathrm{i},u_\mathrm{i}\}$ define the symplectic integrator (e.g. \citealt{Ruth1983, Hairer2006}). A symplectic integrator of any even order exists for every separable Hamiltonian $H=T+U$ and it is possible to find the integrator coefficients $\{t_\mathrm{i},u_\mathrm{i}\}$ efficiently \citep{Yoshida1990,Yoshida1993}.

The integrator coefficients are obtained by using the so-called Campbell-Baker-Hausdorff (hereafter CBH) formula \citep{Campbell1896,Campbell1897,Baker1902,Baker1905,Hausdorff1906}. The CBH formula formally solves the operator $\op{Z}$ from the equation
\begin{equation}
e^{\op{X}+\op{Y}} = e^{\op{Z}}
\end{equation}
as a series expansion of increasingly complex nested commutator expressions. The first few terms of the solution for $\op{Z}$ are
\begin{equation}\label{eq: bch-math}
\begin{split}
\op{Z} &= \log{\left(e^{\op{X}} e^{\op{Y}} \right)}\\
&= \op{X}+\op{Y}+\frac{1}{2}[\op{X},\op{Y}]+\frac{1}{12} \Big( [\op{X},[\op{X},\op{Y}]] + [\op{Y},[\op{Y},\op{X}]]\Big) +  \ldots
\end{split}
\end{equation}
in which $[\op{X},\op{Y}] = \op{XY}-\op{YX}$ is the commutator of operators $\op{X}$ and $\op{Y}$. Inserting Eq. \eqref{eq: Hamiltonian_product_1} into the CBH formula yields the expression
\begin{equation}\label{eq: BCH}
\begin{split}
&\log{\left( \prod_\mathrm{i=1}^\mathrm{N} e^{\epsilon t_\mathrm{i} \op{T}} e^{\epsilon u_\mathrm{i} \op{U}}\right)} = \epsilon \Big( e_\mathrm{T} \op{T} + e_\mathrm{U} \op{U} + \epsilon e_\mathrm{TU} [\op{T},\op{U}] \\ &\phantom{=}\, + \epsilon^2 \left( e_\mathrm{TTU} [\op{T},[\op{T},\op{U}]] + e_\mathrm{UTU} [\op{U},[\op{T},\op{U}]]\,\right) + \ldots\, \Big)\\
&\phantom{=}\, = \epsilon ( \op{H} + \op{H}_\mathrm{err}(\epsilon)) = \epsilon \op{\tilde{H}}.
\end{split}
\end{equation}
The equation reveals the reason for the oscillatory behaviour of the total energy i.e. the Hamiltonian $H$ in symplectic integrators. It originates from the dynamics generated by the error Hamiltonian $H_\mathrm{err}$ \citep{Chin2007a}. The constants $e_\mathrm{T}$, $e_\mathrm{U}$, $e_\mathrm{TU}$, $e_\mathrm{TTU}$ and $e_\mathrm{UTU}$ are the error coefficients of the integrator. They can be computed from the integrator coefficients $\{t_\mathrm{i}\}$ and $\{u_\mathrm{i}\}$ with the constraints  
\begin{equation}
\begin{split}
    &e_\mathrm{T} = \sum_\mathrm{i=1}^\mathrm{N} t_\mathrm{i} = 1\\
    &e_\mathrm{U} = \sum_\mathrm{i=1}^\mathrm{N} u_\mathrm{i} = 1,
\end{split}
\end{equation}
i.e. the coefficients in the drift and kick operators must sum to unity to be consistent with the original time evolution operator. We immediately recognise the familiar first-order Euler integrators
\begin{equation}
\begin{split}
&e^{\epsilon \op{H}} = e^{\epsilon \op{T}} e^{\epsilon \op{U}}\\
&e^{\epsilon \op{H}} = e^{\epsilon \op{U}} e^{\epsilon \op{T}}
\end{split}
\end{equation}
in which the drift and kick operators simply alternate. The error terms
\begin{equation}\label{eq: herr-euler}
\begin{split}
&\op{H}_\mathrm{err}^\mathrm{DK}(\epsilon) = +\frac{1}{2} \epsilon [\op{T},\op{U}] + \ldots\\
&\op{H}_\mathrm{err}^\mathrm{KD}(\epsilon) = -\frac{1}{2} \epsilon [\op{T},\op{U}] + \ldots
\end{split}
\end{equation}
are of the first order as expected.

The most simple generalisation of the first-order Euler integrator is obtained by setting $\{t_\mathrm{1},t_\mathrm{2}\} = \{\sfrac{1}{2},\sfrac{1}{2}\}$, $\{u_\mathrm{1},u_\mathrm{2}\} = \{1,0\}$ or $\{t_\mathrm{1},t_\mathrm{2}\} = \{1,0\}$, $\{u_\mathrm{1},u_\mathrm{2}\} = \{\sfrac{1}{2},\sfrac{1}{2}\}$. We use the BCH formula of Eq. \eqref{eq: bch-math} twice to find that
\begin{equation}
\log{\left(e^{\frac{1}{2} \op{X}} e^{\op{Y}} e^{\frac{1}{2} \op{X}} \right)} 
= \op{X}+\op{Y}-\frac{1}{24} [\op{X},[\op{X},\op{Y}]] + \frac{1}{12} [\op{Y},[\op{Y},\op{X}]] + \ldots
\end{equation}
i.e. the first-order error term has vanished. In fact all the odd terms vanish for all symmetric operator products \citep{Chin2007}. Inserting Eq. \eqref{eq: BCH} into this result yields the common second-order kick-drift-kick
\begin{equation}\label{eq: kdk}
\begin{split}
\log{\left( e^{\frac{1}{2} \op{U}} e^{\op{T}} e^{\frac{1}{2} \op{U}} \right)}= \epsilon \Big( \op{T}+\op{U} - \frac{\epsilon^2}{24} [\op{T},[\op{T},\op{U}]]  {+\frac{\epsilon^2}{12} [\op{U},[\op{T},\op{U}]] + \ldots\,\Big)}
\end{split}
\end{equation}
and drift-kick-drift
\begin{equation}\label{eq: dkd}
\begin{split}
\log{\left( e^{\frac{1}{2} \op{T}} e^{\op{U}} e^{\frac{1}{2} \op{T}} \right)} = \epsilon \Big( \op{T}+\op{U} - \frac{\epsilon^2}{24} [\op{U},[\op{U},\op{T}]] {+ \frac{\epsilon^2}{12} [\op{T},[\op{U},\op{T}]] + \ldots\,\Big)}
\end{split}
\end{equation}
leapfrog integrators. The leading-order error terms generated by the leapfrog error Hamiltonians
\begin{equation}\label{eq: herr-leapfrog}
\begin{split}
& \op{H}^{\mathrm{KDK}}_\mathrm{err}(\epsilon) = - \frac{\epsilon^2}{24} [\op{T},[\op{T},\op{U}]] + \frac{\epsilon^2}{12} [\op{U},[\op{T},\op{U}]] + \ldots\\
& \op{H}^{\mathrm{DKD}}_\mathrm{err}(\epsilon) = - \frac{\epsilon^2}{12} [\op{T},[\op{T},\op{U}]] + \frac{\epsilon^2}{24} [\op{U},[\op{T},\op{U}]] + \ldots
\end{split}    
\end{equation}
are of the second order as expected.

In Euler and leapfrog integrators above the non-zero integrator coefficients $\{t_\mathrm{i},u_\mathrm{i}\}$ are always positive. However, no rule guarantees that the higher-order non-zero integrator coefficients should remain strictly positive \citep{Yoshida1990}. Indeed, it was proven by \cite{Sheng1989} and \cite{Suzuki1991} that beyond second order some of the coefficients $\{t_\mathrm{i},u_\mathrm{i}\}$ must be negative, leading to negative time-steps during the integration. In addition, \cite{Goldman1996} found that both $\{t_\mathrm{i}\}$ and $\{u_\mathrm{i}\}$ must contain at least a single negative coefficient. In general negative time-steps prohibit the integration of time-irreversible systems such as ones with dissipation (e.g. \citealt{Chin2007}) and can make hierarchical symplectic integration schemes inefficient \citep{Pelupessy2012}.

\subsection{Fourth-order forward symplectic integrators}\label{section: fsi}

The essence of the solution to the issue of the negative integration time-steps in high-order symplectic integrators can be understood by studying the first terms of the leapfrog error Hamiltonian. The key idea is to move one of the double commutators in Eq. \eqref{eq: herr-leapfrog} into the actual Hamiltonian instead of including it in the error Hamiltonian as before. This allows setting the remaining error coefficient, which is either $e_\mathrm{TTU}$ or $e_\mathrm{UTU}$, to zero \citep{Chin2005, Chin2007a, Dehnen2017a}. If one keeps the operator product symmetric then the leading-order error terms are of the fourth order and the integrator coefficients are strictly positive. 

The next question is to decide which double commutator (either $[\op{T},[\op{T},\op{U}]]$ or $[\op{U},[\op{T},\op{U}]]$) in Eq. \eqref{eq: herr-leapfrog} is to be moved into the Hamiltonian and which one is to be discarded. The solution is to set $e_\mathrm{TTU} = 0$ and move the double commutator $[\op{U},[\op{T},\op{U}]]$ into the Hamiltonian. This is because $[\op{U},[\op{T},\op{U}]]$ can be shown to correspond to a calculable scalar function which only depends on the coordinates of the dynamical system \citep{Takahashi1984}, so it corresponds to an extra potential term $G$ \citep{Dehnen2017a} in the Hamiltonian defined as
\begin{equation}\label{eq: potential-gradient}
    [U,[T,U]] = -\sum_\mathrm{i}^\mathrm{N} \frac{1}{m_\mathrm{i}} \left( \pderfrac{U}{q_\mathrm{i}} \right)^2 = -\sum_\mathrm{i}^\mathrm{N} m_\mathrm{i} \norm{\vect{a}_\mathrm{i}}^2 \equiv G.
\end{equation}

Now we are ready to perform the actual derivation of the fourth-order forward integrator. We begin from the symmetric operator product relation 
\begin{equation}\label{eq: bch-3}
\log{\left(e^{\frac{1}{6} \op{X}} e^{\frac{1}{2} \op{Y}} e^{\frac{2}{3} \op{X}} e^{\frac{1}{2} \op{Y}} e^{\frac{1}{6} \op{X}} \right)} 
= \op{X}+\op{Y}-\frac{1}{72} [\op{X},[\op{Y},\op{X}]] + \ldots
\end{equation}
which can be derived by using the BCH formula three times \citep{Dehnen2017a}. The formula indicates that the corresponding integrator has the error Hamiltonian of 
\begin{equation}
\op{H}_\mathrm{err}^{\mathrm{KDKDK}} = - \frac{1}{72} [\op{U},[\op{T},\op{U}]] + \ldots
\end{equation}
which we already know to be calculable. The term should be placed in the operator product of the integrator in such a way that the product remains symmetric. From the computational point of view the he optimal location is within in the term in the middle of Eq. \eqref{eq: bch-3} to avoid evaluating the term more than once. The Hamiltonian which generates the dynamics of the fourth-order forward integrator is
\begin{equation}
H = T + U + \frac{1}{48} [U,[T,U]]  = T+U+\frac{1}{48} \epsilon^2 G
\end{equation}
with the leading term of the error Hamiltonian being $H_\mathrm{err} = \bigO{\epsilon^4}$.
The time evolution operator for this Hamiltonian is 
\begin{equation}\label{eq: fsi}
\begin{split}
e^{\epsilon \op{H}} &= e^{\frac{1}{6} \epsilon \op{U}} e^{\frac{1}{2} \epsilon \op{T}} e^{\frac{2}{3} \epsilon \left( \op{U} + \frac{1}{48} \epsilon^2 \op{G} \right) } e^{\frac{1}{2} \epsilon \op{T}} e^{\frac{1}{6} \epsilon \op{U}}\\
&= e^{\frac{1}{6} \epsilon \op{U}} e^{\frac{1}{2} \epsilon \op{T}} e^{\frac{2}{3} \epsilon \op{\tilde{U}} } e^{\frac{1}{2} \epsilon \op{T}} e^{\frac{1}{6} \epsilon \op{U}}
\end{split}
\end{equation}
in which $\op{\tilde{U}}$ corresponds to the so-called modified or gradient potential defined as
\begin{equation}\label{eq: mod-pot}
    \tilde{U} = U + \frac{1}{48} \epsilon^2 G.
\end{equation}
The integrator Eq.~\eqref{eq: fsi} is known as the forward symplectic integrator or FSI \citep{Chin2007} or the gradient symplectic integrator especially in the early literature. The FSI presented here is only a single example of the class of fourth-order symplectic integrators which were found and studied by \cite{Chin1997} and \cite{Chin2005} based on the pioneering work of \cite{Ruth1983, Takahashi1984} and \cite{Suzuki1995}.

\subsection{Gradient force expressions for direct N-body codes}

Next we turn into practical matters show how the FSI can be numerically implemented into a direct N-body code. This is a rather straightforward task as the only new expression to be calculated is the formula for the so-called gradient force $\vect{\tilde{F}}_\mathrm{i}$ (or acceleration $\vect{\tilde{a}}_\mathrm{i}$) which originates from the modified potential $\tilde{U}$ in Eq. \eqref{eq: mod-pot}.

The Hamiltonian for an N-body system in Newtonian gravity is defined as
\begin{equation}
\begin{split}
H = T + U &= \frac{1}{2} \sum_\mathrm{i=1}^\mathrm{N} \frac{ \norm{\vect{p_\mathrm{i}}}^2 }{m_\mathrm{i}} - \mathcal{G} \sum_\mathrm{i=1}^\mathrm{N} \sum_\mathrm{j>i}^\mathrm{N} \frac{m_\mathrm{i} m_\mathrm{j}}{\norm{\vect{x}_\mathrm{ji}}}\\ &= \frac{1}{2} \sum_\mathrm{i=1}^\mathrm{N} m_\mathrm{i} \norm{\vect{v_\mathrm{i}}}^2 - \mathcal{G} \sum_\mathrm{i=1}^\mathrm{N} \sum_\mathrm{j>i}^\mathrm{N} \frac{ m_\mathrm{i} m_\mathrm{j}}{r_\mathrm{ji}}
\end{split}
\end{equation}
in which we switched into somewhat more relaxed notation. The separation vectors and their norms here are defined as $\vect{x}_\mathrm{ji} = \vect{x}_\mathrm{j} - \vect{x}_\mathrm{i}$ and $r_\mathrm{ji} = \norm{\vect{x}_\mathrm{ji}}$ and the individual particle masses $m_\mathrm{i}$ are constant. Here $\mathcal{G}$ is Newton's constant.

In a system of $N$ bodies the Newtonian acceleration of a body is computed as \begin{equation}\label{eq: acc-newton}
\vect{a}_\mathrm{i} = - \frac{1}{m_\mathrm{i}}\pderfrac{U}{x_\mathrm{i}} = \mathcal{G} \sum_\mathrm{j \neq i}^\mathrm{N} m_\mathrm{j} \frac{\vect{x}_\mathrm{ji}}{r_\mathrm{ji}^3}.
\end{equation}
The expression for acceleration corresponding to potential $G$ of Eq. \eqref{eq: potential-gradient} is somewhat more complicated. We begin by calculating the gradient acceleration $\vect{g_\mathrm{i}}$ for a single test particle with mass $m_\mathrm{i}$ in a gradient potential of a massive body $M$ located fixed at the origin. The result is
\begin{equation}
\begin{split}
\vect{g}_\mathrm{i} &= -\frac{1}{m_\mathrm{i}}\pderfrac{G}{\vect{x}_\mathrm{i}} 
= \pderfrac{}{\vect{x}_\mathrm{i}} \norm{\vect{a}_\mathrm{i}}^2 = 2 \pderfrac{a_\mathrm{i}}{\vect{x}_\mathrm{i}} \cdot \vect{a}_\mathrm{i}\\ &= 2 \frac{\mathcal{G}M}{r^5_\mathrm{i}} \bigg( r_\mathrm{i}^2 \vect{a}_\mathrm{i} - 3 ( \vect{x}_\mathrm{i} \cdot \vect{a}_\mathrm{i} ) \vect{x}_\mathrm{i} \bigg) = 2\, t_\mathrm{dyn}^{-2}\, \vect{a}_\mathrm{i}.
\end{split}
\end{equation}
in which $t_\mathrm{dyn}$ is the dynamical timescale of the test particle. Thus, the total potential in Eq. \eqref{eq: potential-gradient} generates the following acceleration for the test particle: 
\begin{equation}
    \tilde{\vect{a}_\mathrm{i}} = \vect{a}_\mathrm{i} + \frac{1}{48} \epsilon^2 \vect{g}_\mathrm{i} = \left[1 +  \frac{1}{24} \left( \frac{\epsilon}{t_\mathrm{dyn}}\right)^2 \right] \vect{a}_\mathrm{i}.
\end{equation}
We note that the expression closely resembles \citep{Chin2007} the extrapolated effective gradient force of \cite{Omelyan2006}.

The N-body case is a straightforward generalisation of the test particle scenario. The main difference is the use of relative accelerations $\vect{a}_\mathrm{ji} = \vect{a}_\mathrm{j}-\vect{a}_\mathrm{i}$ in the formulas:
\begin{equation}\label{eq: acc-gradient-g}
\begin{split}
\vect{g}_\mathrm{i} &= -\frac{1}{m_\mathrm{i}}\pderfrac{G}{\vect{x}_\mathrm{i}} 
= \pderfrac{}{\vect{x}_\mathrm{i}} \sum_\mathrm{j}^\mathrm{N} \norm{\vect{a}_\mathrm{j}}^2\\ &= 2 \sum_\mathrm{j \neq i}^\mathrm{N} \frac{\mathcal{G} m_\mathrm{j}}{r^5_\mathrm{ji}} \bigg( r_\mathrm{ji}^2 \vect{a}_\mathrm{ji} - 3 ( \vect{x}_\mathrm{ji} \cdot \vect{a}_\mathrm{ji} ) \vect{x}_\mathrm{ji} \bigg).
\end{split}
\end{equation}
The modified potential $\tilde{U}$ generates N-body accelerations $\tilde{\vect{a}_\mathrm{i}}$ of
\begin{equation}\label{eq: acc-gradient}
\begin{split}
\tilde{\vect{a_\mathrm{i}}} & = \vect{a}_\mathrm{i} + \frac{1}{48} \epsilon^2 \vect{g}_\mathrm{i}\\
& = \vect{a}_\mathrm{i} + \frac{\mathcal{G} \epsilon^2}{24} \sum_\mathrm{j \neq i}^\mathrm{N} \frac{m_\mathrm{j}}{r^5_\mathrm{ji}} \bigg( r_\mathrm{ji}^2 \vect{a}_\mathrm{ji} - 3 ( \vect{x}_\mathrm{ji} \cdot \vect{a}_\mathrm{ji} ) \vect{x}_\mathrm{ji} \bigg)
\end{split}
\end{equation}
in which again $\vect{a}_\mathrm{ji} = \vect{a}_\mathrm{j}-\vect{a}_\mathrm{i}$. Note that computing the gradient accelerations requires a second sum over the particles whereas the in the case of the Newtonian acceleration only one sum is needed.

Gravitational softening (e.g. \citealt{Barnes2012}) may be included as well. If so, one has to replace the potential $U$ with a softened one in Eq. \eqref{eq: acc-newton} and Eq. \eqref{eq: acc-gradient} and compute the two softened accelerations. For example the common Plummer softening kernel \citep{Plummer1911} can be included in a straightforward manner by substituting $r_\mathrm{ji}$ with $(r_\mathrm{ji}^2 + \epsilon_\mathrm{P}^2)^{1/2}$ in which $\epsilon_\mathrm{P}$ is the gravitational softening length. The gravitational softening used in a number of simulations in this study is the Plummer softening.


\section{Symplectic integrators from hierarchical Hamiltonian splitting}\label{section: 3}

\subsection{Hierarchical second-order integrators}

Hierarchical Hamiltonian splitting (hereafter HHS) is a technique to construct symplectic N-body integrators with individual time-steps for the particles of the dynamical system \citep{Pelupessy2012}. The advantages of well-constructed HHS integrators compared to integrators with common block time-steps are manifest momentum conservation and extremely large dynamical range. In this Section we generalise the second-order hierarchical symplectic integrator of \cite{Pelupessy2012} into a hierarchical fourth-order integrator with strictly positive time-steps.

The key idea of the HHS scheme is to first assign individual time-steps to particles and then to divide the particles into two sets of so-called slow and fast particles using a pivot time-step. The slow particles are then propagated using the pivot time-step while the fast particles are divided again now using half of the pivot time-step and so on. This process is applied recursively until all the particles have been propagated.

More rigorously, given an initial pivot time-step $\tau_\mathrm{pivot,1}$, corresponding to the maximum particle time-step in the block time-step scheme, the particles $\mathcal{P}_\mathrm{i}$ of an N-body system $\{\mathcal{P}_\mathrm{i}\}$ are divided into two non-overlapping sets of slow $\mathcal{S}_\mathrm{j}$ and fast $\mathcal{F}$ particles. In general, the subscript $j$ labels the hierarchy level of the pivot step as $\tau_\mathrm{pivot,j}$. The division criterion is based on the individual time-steps $\tau_\mathrm{i}$ of the particles as
\begin{equation}
\begin{split}
&\Big\{\,\mathcal{S}_\mathrm{j} = \big\{\mathcal{P}_\mathrm{i} \big\}\;\; \Big|\; \tau_\mathrm{i} \geq \tau_\mathrm{pivot,j}\, \Big\}\hspace{1.0cm}\text{set of slow particles}\\
&\Big\{\,\mathcal{F}\, = \big\{\mathcal{P}_\mathrm{k} \big\}\; \Big|\; \tau_\mathrm{k} < \tau_\mathrm{pivot,j}\, \Big\}\hspace{0.97cm}\text{set of fast particles}\\
\end{split}
\end{equation}
with $\mathcal{S}_\mathrm{j} \cup \mathcal{F}$ being  equal to the original particle set. When $\mathcal{F}$ is further partitioned the pivot new step is $\tau_\mathrm{pivot,j+1} = \frac{1}{2}\tau_\mathrm{pivot,j}$ and the time-steps $\tau_\mathrm{i}$ are re-computed taking only the particles in $\mathcal{F}$ into account. The Hamiltonian of the particle system at each level of time-step hierarchy is split according to the two sets $\mathcal{S}$ and $\mathcal{F}$ as
\begin{equation}\label{eq: fast-slow-split}
    H = H_\mathrm{S} + H_\mathrm{F} + H_\mathrm{SF}.
\end{equation}
Here $H_\mathrm{S}=T_\mathrm{S} +U_\mathrm{SS}$ and $H_\mathrm{F}=T_\mathrm{F} + U_\mathrm{FF}$ are the Hamiltonians of the sets of slow $\mathcal{S}$ and fast particles $\mathcal{F}$ on the particular hierarchy level. The third term $H_\mathrm{SF} = U_\mathrm{SF} = U_\mathrm{FS}$ is the interaction Hamiltonian between the two systems which guarantees that acceleration calculations and kick operations between particles on different levels of hierarchy are always pair-wise i.e. synchronised. This is the origin of the manifest momentum conservation of the HHS integrators. After the procedure has been recursively repeated on $\mathcal{F}$ until no particles remain, we are left with a collection of slow Hamiltonians of the sets $\mathcal{S}_\mathrm{j}$ and their mutual interaction Hamiltonians.

The Hamiltonian of Eq. \eqref{eq: fast-slow-split} can be used to generate various time evolution operators for practical integration algorithms. \cite{Pelupessy2012} studied a number of these integrators in second order and found that the following so-called HOLD algorithm has the best numerical performance. The time evolution operator of the HOLD integrator is derived from the Hamiltonian of Eq. \eqref{eq: fast-slow-split} as
\begin{equation}\label{eq: hhs2}
\begin{split}
    &e^{\epsilon \op{H}} = e^{\epsilon (\op{H}_\mathrm{F} + \op{H}_\mathrm{S} + \op{H}_\mathrm{FS})} =e^{\epsilon (\op{H}_\mathrm{F} + \op{H}_\mathrm{S} + \op{U}_\mathrm{FS}) } = e^{\epsilon (\op{H}_\mathrm{F} + \op{H}_\mathrm{S}) + \epsilon \op{U}_\mathrm{FS}) }\\
    &\approx e^{\frac{1}{2} \epsilon \left( \op{H}_\mathrm{F}+\op{H}_\mathrm{S}\right) } e^{ \epsilon \op{U}_\mathrm{FS} } e^{\frac{1}{2} \epsilon \left(\op{H}_\mathrm{F}+\op{H}_\mathrm{S}\right)}\\
    &=e^{ \frac{1}{2} \epsilon \op{H}_\mathrm{F} } e^{\frac{1}{2} \epsilon \op{H}_\mathrm{S}} e^{ \epsilon \op{U}_\mathrm{FS} } e^{\frac{1}{2} \epsilon \op{H}_\mathrm{S}} e^{\frac{1}{2} \epsilon \op{H}_\mathrm{F}}.
\end{split}
\end{equation}
The final step follows from the fact that time evolution operators generated by $H_\mathrm{S}$ and $H_\mathrm{F}$ commute by definition. 

The name HOLD of the integrator originates from the notion that it is advantageous to keep, or hold, the slow-fast interaction term at the slow level of the hierarchy \citep{Pelupessy2012}. This fact has formidable consequences: the Hamiltonian of a certain level in the time-step hierarchy is independent of the slower hierarchy levels. Computationally this implies that one can efficiently focus on the internal dynamics of particle systems with very short time-steps ignoring the particles with longer time-steps. In the block time-step schemes the force calculation of the active particles with small time-steps requires taking every particle of the entire dynamical system into account. In addition, all the interactions between the different levels in the time-step hierarchy are again always pair-wise so the HOLD integration algorithm is manifestly momentum-conserving.

Finally, the time evolution operator for the Hamiltonian of the slow particle set $H_\mathrm{S}$ in Eq. \eqref{eq: hhs2} is the common second-order leapfrog integrator i.e.
\begin{equation}
e^{\epsilon \op{H}_\mathrm{S}} = e^{\epsilon ( \op{T}_\mathrm{S} + \op{U}_\mathrm{SS} ) } \approx e^{\frac{1}{2} \epsilon \op{U}_\mathrm{SS} } e^{\epsilon \op{T}_\mathrm{S}} e^{\frac{1}{2} \epsilon \op{U}_\mathrm{SS}}.
\end{equation}
In principle nothing prohibits using higher-order symplectic integrators for $H_\mathrm{S}$ for improved integration accuracy. However, using e.g. a high-order Yoshida integrator would not change the order of the HOLD integration method as the initial Hamiltonian splitting into slow and fast Hamiltonians in Eq. \eqref{eq: fast-slow-split} was of the second order.

\subsection{A new hierarchical fourth-order forward symplectic integrator}

\begin{figure*}
\includegraphics[width=\textwidth]{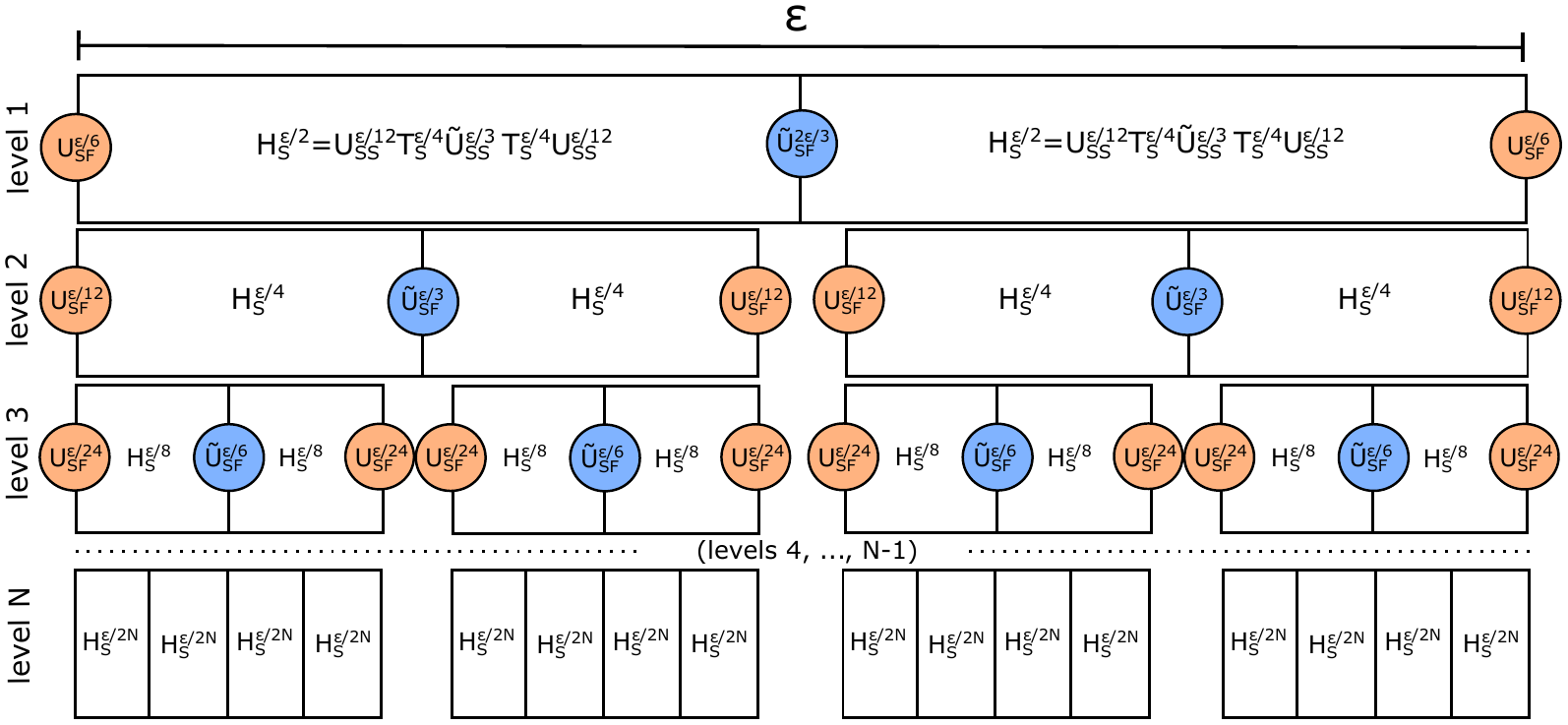}
\caption{A time-step hierarchy chart of the fourth-order HHS-FSI integrator with $N$ hierarchical levels. Here $H$ represents the time evolution operator of Eq. \eqref{eq: time-evolution-operator-hhs-fsi} while $T$ and $U$ are the drift and kick operators. The subscripts $S$ and $F$ label the systems of slow and fast particles. The time-step of the slowest hierarchy level is $\epsilon$.}
\label{fig-flowchart}
\end{figure*}
We now construct a novel hierarchical symplectic fourth-order integration algorithm HHS-FSI which has strictly positive time-steps. First we must split the time evolution operator generated by the Hamiltonian of Eq. \eqref{eq: fast-slow-split} i.e.
\begin{equation}\label{eq: fast-slow-H}
e^{\epsilon \op{H}} = e^{\epsilon \left(\op{H}_\mathrm{S} + \op{H}_\mathrm{F} + \op{H}_\mathrm{SF}\right)} = e^{\epsilon
\left(\op{H}_\mathrm{S} + \op{H}_\mathrm{F} + \op{U}_\mathrm{SF}\right)} = e^{\epsilon
(\op{H}_\mathrm{S} + \op{H}_\mathrm{F} ) + \epsilon \op{U}_\mathrm{SF}}
\end{equation}
using a fourth-order splitting scheme following the recipe presented in Section $\ref{section: fsi}$. There are two symmetric possibilities how to do this. One may either place the operators generated by interaction Hamiltonian $U_\mathrm{SF}$ in the middle and both ends of the operator product as
\begin{equation}\label{eq: 4th-order-split-possiblity-1}
e^{\epsilon \op{H}} = e^{\frac{1}{6} \epsilon \op{U}_\mathrm{SF}} e^{\frac{1}{2} \epsilon \left( \op{H}_\mathrm{S} + \op{H}_\mathrm{F}\right)} e^{\frac{2}{3} \epsilon \op{U}_\mathrm{SF}} e^{\frac{1}{2} \epsilon \left(\op{H}_\mathrm{S} + \op{H}_\mathrm{F}\right)}e^{\frac{1}{6} \epsilon \op{U}_\mathrm{SF}}    
\end{equation}
or set the operators generated by $H_\mathrm{S}$ and $H_\mathrm{F}$ into these locations i.e.
\begin{equation}\label{eq: 4th-order-split-possiblity-2}
e^{\epsilon \op{H}} = e^{\frac{1}{6} \epsilon ( \op{H}_\mathrm{S} + \op{H}_\mathrm{F} ) } e^{\frac{1}{2} \epsilon \op{U}_\mathrm{SF} } e^{\frac{2}{3} \epsilon ( \op{H}_\mathrm{S} + \op{H}_\mathrm{F} )  } e^{\frac{1}{2} \epsilon \op{U}_\mathrm{SF} } e^{\frac{1}{6} \epsilon ( \op{H}_\mathrm{S} + \op{H}_\mathrm{F} ) }.
\end{equation}
The two error Hamiltonians of these time evolution operators are
\begin{align}
&\op{H}_\mathrm{err} = - \frac{1}{72}  [\op{U}_\mathrm{SF},[\op{H}_\mathrm{S}+\op{H}_\mathrm{F},\op{U}_\mathrm{SF}] + \ldots\\
&\op{H}_\mathrm{err} = - \frac{1}{72} [\op{H}_\mathrm{S}+\op{H}_\mathrm{F},[\op{U}_\mathrm{SF},\op{H}_\mathrm{S}+\op{H}_\mathrm{F}]] + \ldots
\end{align}
by the BCH identity of Eq. \eqref{eq: bch-3}.

To avoid negative integrator coefficients following Section \ref{section: fsi} we must evaluate one of the error double commutators and discard the other. Calculating the two commutator expressions we find that the latter double commutator is not suitable for our purposes as it results in a non-separable Hamiltonian. Thus, the time evolution operator of Eq. \eqref{eq: 4th-order-split-possiblity-2} does not correspond to a practical fourth-order forward integration algorithm. Fortunately, the former double commutator can be evaluated as
\begin{equation}\label{eq: tut-utu}
\begin{split}
&[\op{U}_\mathrm{SF},[\op{H}_\mathrm{S}+H_\mathrm{F},\op{U}_\mathrm{SF}]] = [\op{U}_\mathrm{SF},[\op{T}_\mathrm{S}+\op{T}_\mathrm{F},\op{U}_\mathrm{SF}]]\\ 
&= [\op{U}_\mathrm{SF},[\op{T}_\mathrm{S},\op{U}_\mathrm{SF}]] + [\op{U}_\mathrm{SF},[\op{T}_\mathrm{F},\op{U}_\mathrm{SF}]].
\end{split}
\end{equation}
so Eq. \eqref{eq: 4th-order-split-possiblity-1} yields the integrator we search for. The two terms of Eq. \eqref{eq: tut-utu} have an interpretation analogous to Eq. \eqref{eq: potential-gradient}: they correspond to gradient potentials of the sets of slow and fast particle sets as 
\begin{equation}\label{eq: gs-gf}
\begin{split}
&[U_\mathrm{SF},[T_\mathrm{S},U_\mathrm{SF}]] = -\sum_\mathrm{i\in \mathcal{S}} \frac{1}{m_\mathrm{i}} \left( \pderfrac{U_\mathrm{SF}}{x_\mathrm{i}} \right)^2 \equiv G_\mathrm{S}\\
&[U_\mathrm{SF},[T_\mathrm{F},U_\mathrm{SF}]] = -\sum_\mathrm{i\in \mathcal{F}} \frac{1}{m_\mathrm{i}} \left( \pderfrac{U_\mathrm{SF}}{x_\mathrm{i}} \right)^2 \equiv G_\mathrm{F}.
\end{split}
\end{equation}
The Hamiltonian generating the dynamics of the HHS-FSI integrator is thus
\begin{equation}
\begin{split}
H &= H_\mathrm{S} + H_\mathrm{F} + \frac{1}{48} \epsilon^2 \left( G_\mathrm{S} + G_\mathrm{F}\right)\\
&=H_\mathrm{S} + H_\mathrm{F} + U_\mathrm{SF} + \tilde{U}_\mathrm{SF}.
\end{split}
\end{equation}
The corresponding time evolution operator $e^{\epsilon \op{H}}$ is
\begin{equation}\label{eq: time-evolution-operator-hhs-fsi}
\begin{split}
&e^{\frac{1}{6} \epsilon \op{U}_\mathrm{SF}} e^{\frac{1}{2} \epsilon \left( \op{H}_\mathrm{S} + \op{H}_\mathrm{F}\right)} e^{\frac{2}{3} \epsilon \left( \op{U}_\mathrm{SF} + \frac{1}{48} \epsilon^2 (\op{G}_\mathrm{S} + \op{G}_\mathrm{F}) \right)} e^{\frac{1}{2} \epsilon \left(\op{H}_\mathrm{S} + \op{H}_\mathrm{F}\right)}e^{\frac{1}{6} \epsilon \op{U}_\mathrm{SF}}\\
&= e^{\frac{1}{6} \epsilon \op{U}_\mathrm{SF}} e^{\frac{1}{2} \epsilon \op{H}_\mathrm{S} } e^{\frac{1}{2} \epsilon \op{H}_\mathrm{F} } e^{\frac{2}{3} \epsilon \op{\tilde{U}}_\mathrm{SF}} e^{\frac{1}{2} \epsilon \op{H}_\mathrm{F} } e^{\frac{1}{2} \epsilon \op{H}_\mathrm{S} }e^{\frac{1}{6} \epsilon \op{U}_\mathrm{SF}} ,
\end{split}
\end{equation}
in which we have again used the fact that the time evolution operators generated by the slow and fast Hamiltonians commute. The HHS-FSI integrator of Eq. \eqref{eq: time-evolution-operator-hhs-fsi} is the main result of this study and it is implemented in the novel direct N-body code \frost{} in Section \ref{section: cuda}.

The systems of slow particles with Hamiltonian $H_\mathrm{S}$ are propagated using the FSI integrator from Section \ref{section: fsi}. Here it is possible to use other symplectic integrators of at least of the fourth order \citep{Mclachlan1995} without decreasing the order of the entire HHS-FSI integration algorithm.

\subsection{Gradient force expressions between time-step hierarchy levels for direct N-body codes}

The gradient accelerations $\tilde{\vect{a}}_\mathrm{i}$ corresponding to the gradient potentials $G_\mathrm{S}$ and $G_\mathrm{S}$ between the sets of slow and fast particles ($\mathcal{S}$ and $\mathcal{F}$) in Eq. \eqref{eq: gs-gf} can be computed starting from the corresponding Newtonian acceleration formulas for the particles $\mathcal{P}_\mathrm{i}$ and $\mathcal{P}_\mathrm{j}$ as
\begin{equation}\label{eq: force-newton-inter}
\begin{split}
&\vect{a}_\mathrm{i} = \mathcal{G} \sum_\mathrm{j \in \mathcal{F}} m_\mathrm{j} \frac{ \vect{x}_\mathrm{ji}}{r_\mathrm{ji}^3} \hspace{1cm} \text{if $\mathcal{P}_\mathrm{i} \in \mathcal{S}$}\\
&\vect{a}_\mathrm{j} = \mathcal{G} \sum_\mathrm{i\in \mathcal{S}} m_\mathrm{i} \frac{\vect{x}_\mathrm{ij}}{r_\mathrm{ij}^3} \hspace{1cm} \text{if $\mathcal{P}_\mathrm{j} \in \mathcal{F}$}.
\end{split}
\end{equation}
The gradient accelerations between the two time-step hierarchy levels are thus
\begin{equation}\label{eq: force-gradient-inter-1}
\begin{split}
\tilde{\vect{a}_\mathrm{i}} & = \vect{a}_\mathrm{i} + \frac{1}{48} \epsilon^2 \vect{g}_\mathrm{i}\\ &= \vect{a}_\mathrm{i} + \frac{\mathcal{G} \epsilon^2}{24} \sum_\mathrm{j \in \mathcal{F}} \frac{ m_\mathrm{j}}{r^5_\mathrm{ji}} \bigg( r_\mathrm{ji}^2 \vect{a}_\mathrm{ji} - 3 (\vect{x}_\mathrm{ji} \cdot \vect{a}_\mathrm{ji} ) \vect{x}_\mathrm{ji} \bigg) \hspace{0.3cm }\text{if $\mathcal{P}_\mathrm{i} \in \mathcal{S}$}
\end{split}
\end{equation}
and 
\begin{equation}\label{eq: force-gradient-inter-2}
\begin{split}
\tilde{\vect{a}_\mathrm{j}} & = \vect{a}_\mathrm{j} + \frac{1}{48} \epsilon^2 \vect{g}_\mathrm{j}\\ &= \vect{a}_\mathrm{j} + \frac{\mathcal{G} \epsilon^2}{24} \sum_\mathrm{i \in \mathcal{S}} \frac{m_\mathrm{i}}{r^5_\mathrm{ij}} \bigg( r_\mathrm{ij}^2 \vect{a}_\mathrm{ij} - 3 ( \vect{x}_\mathrm{ij} \cdot \vect{a}_\mathrm{ij} ) \vect{x}_\mathrm{ij} \bigg) \hspace{0.3cm }\text{if $\mathcal{P}_\mathrm{j} \in \mathcal{F}$}.
\end{split}
\end{equation}


\section{Individual adaptive time-steps and time-step symmetrisation}\label{section: 4}

\subsection{Time-irreversibility of common time-step schemes}

Widely used time-step schemes and time-step criteria typically break the two desirable properties of an integrator: symplecticity and time-reversibility.
In general symplecticity of an integrator already breaks down if the used time-step function $\tau$ depends on the phase-space coordinates $\{\{\vect{x}_\mathrm{i}\}, \{\vect{p}_\mathrm{i}\}\}$ of the system. This occurs as the time evolution operator is not a proper canonical transformation anymore (e.g. \citealt{Dehnen2017b}). In addition, block time-step schemes without mutual pairwise kicks break the symplecticity by rendering the Hamiltonian $H=T(\{\vect{p}_\mathrm{i}\})+U(\{\vect{x}_\mathrm{i}\})$ formally non-separable due to coupling of particles on different time-step blocks \citep{Springel2005}. Our integrator using strictly pair-wise kicks avoids the latter issue \citep{Saha1994,Farr2007,Pelupessy2012}. The loss of symplecticity often leads to secular error growth in the form of numerical dissipation.

Another source of numerical dissipation arises if the time-reversibility of the integrator is broken. A time-symmetric integration recipe loses its time-reversibility if the time-steps depend on the phase-space coordinates $\{\{\vect{x}_\mathrm{i}\}, \{\vect{p}_\mathrm{i}\}\}$ of the system (e.g. \citealt{Preto1999,Pelupessy2012,Hernandez2018}) and the time-step function is evaluated before integrating the time-step. As the time-steps then depend asymmetrically on the past and not the future  phase-space state of the system \citep{Springel2005} the time symmetry is broken. This occurs in most commonly used time-step schemes. Certain special recipes for time-symmetric integration exist (see e.g. Appendix A of \citealt{Hands2019}) but unfortunately not for integrators with discretised (block) time-step schemes \citep{Dehnen2017b}.

\subsection{Time-step symmetrisation}

Numerical methods to mitigate the effects time-irreversibility for integrators using hierarchical or block time-steps have been devised and implemented \citep{Hut1995, Pelupessy2012,AguilarArguello2020}. The time-irreversibility or time synchronisation error can be reduced from $\bigO{\tau}$ to $\bigO{\tau^3}$ \citep{Dehnen2017b} by a method we call in this study the (partial) time-step symmetrisation procedure. The procedure involves extrapolating the time-step functions into the future using their time derivatives before integrating the step.

The time-reversibility of the integration with adaptive time-steps can be summarised in the statement
\begin{equation}
    \tau^{\mathrm{+}}(t) = \tau^{\mathrm{-}}(t+\tau(t))   
\end{equation}
in which the superscript signs indicate the direction of the integration in time. A symmetrised time-step $\tau^\mathrm{sym}$ can be defined e.g. as
\begin{equation}\label{eq: arithmetic}
    \tau^\mathrm{sym} = \frac{1}{2}  \bigg[ \tau(t) +  \tau(t + \tau(t)) \bigg]
\end{equation}
using the common arithmetic mean (as in \citealt{Pelupessy2012}). However, we note that this is not the only possible definition for $\tau^\mathrm{sym}$. The harmonic mean can be used in the symmetrisation formula as well (e.g. \citealt{Holder1999}) yielding the definition
\begin{equation}\label{eq: harmonic}
    \tau^\mathrm{sym} = 2 \left[ \frac{1}{\tau(t)} + \frac{1}{\tau(t + \tau(t))} \right]^{-1}.
\end{equation}
The third obvious option would be to use the geometric mean defined as $\tau^\mathrm{sym} = (\tau(t) \tau(t+\tau(t))^{1/2}$. However, symmetrisation procedures involving products of discretised time-steps are strongly affected by the so-called flip-flop problem \citep{Dehnen2011, Pelupessy2012}. While the problem can be costly circumvented \citep{Makino2006} we only resort to the harmonic and arithmetic means for the rest of this study.

\begin{figure}
\includegraphics[width=\linewidth]{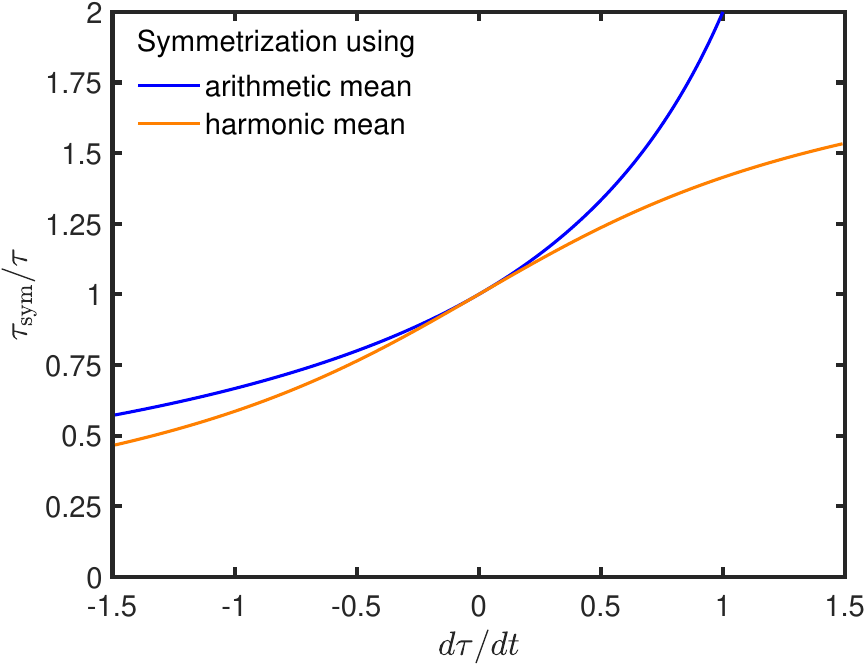}
\caption{The time-step symmetrisation factor $\tau_\mathrm{sym}$/$\tau$ as a function of the time derivative of the time-step $\tau$. Here $\tau_\mathrm{sym}$ is the final symmetrised time-step. The expressions of the arithmetic (blue line) and the symmetric (orange line) time-step factors are shown from Eq. \eqref{eq: time-step-factor}. We use the harmonic time-step symmetrisation in this study for its well-behaving mathematical expression as explained in the text. Note that the symmetrised harmonic time-steps are practically always shorter than the arithmetic time-steps. The two definitions of the symmetrisation factor are identical when the time derivative of the time-step is zero.}
\label{fig: symmetrization}
\end{figure}

Up to this point the definition of the symmetric time-step was exact. In order to proceed towards a practical time-step symmetrisation recipe \citep{Pelupessy2012} we expand $\tau(t+\tau^\mathrm{sym})$ to the first order around the time $t$ as
\begin{equation}
    \tau(t+\tau^\mathrm{sym}) \approx \tau(t) + \tau^\mathrm{sym} \derfrac{\tau(t)}{t}.
\end{equation}
Plugging the low-order expansion into the definition of the arithmetic time-step symmetrisation Eq. \eqref{eq: arithmetic} and the harmonic time-step symmetrisation Eq. \eqref{eq: harmonic} yields two expressions for the symmetrised time-step $\tau^\mathrm{sym}$:
\begin{equation}\label{eq: time-step-factor}
    \begin{aligned}
        \tau^\mathrm{sym}_\mathrm{arithmetic} &= \left[ 1-\frac{1}{2}\derfrac{\tau}{t} \right]^{-1} \tau \hspace{1.55cm} \text{if } \derfrac{\tau}{t} < 2\\
        \tau^\mathrm{sym}_\mathrm{harmonic} &=
        \left[ \derfrac{\tau}{t} + \left( 1+ \left(\derfrac{\tau}{t}\right)^2 \right)^{1/2} -1 \right] \left(\derfrac{\tau}{t} \right)^{-1} \tau
    \end{aligned}
\end{equation}
in which both $\tau$ and its time derivative are evaluated at time $t$.

A few remarks should be noted about both time-step factor expressions above. The arithmetic time-step is physically meaningful (i.e. finite and non-negative) only for time-step derivative values $\ud{\tau}/\ud{t}<2$. A straightforward solution for this is just to limit the maximum value of the derivative. However, we note that if $\tau$ is multiplied by a constant factor its time derivative changes by this factor as well. Thus a more elegant solution is to lower the time-step until is time derivative again fulfils the condition $\ud{\tau}/\ud{t}<2$. 

Concerning the harmonic time-step factor we note that we have discarded a mathematically valid solution during the derivation which would have led into negative or infinite time-step factors. The solution in Eq. \eqref{eq: time-step-factor} selected for this study is continuous and differentiable everywhere, even when the derivative of the time-step is zero. The expression of the symmetrised harmonic time-step is anti-symmetric w.r.t. the point $(0,\tau)$. The expression for the symmetrised arithmetic time-step does not share this property. For these reasons we consider the harmonic time-step symmetrisation factor mathematically somewhat more elegant than the arithmetic factor. In addition, the harmonically symmetrised time-steps are always shorter (or equal) than the arithmetically symmetrised time-steps. This fact originates directly from the definitions of the arithmetic and the harmonic means. The expressions for the two time-step factors $\tau_\mathrm{sym}/\tau$ are visualised in Fig. \ref{fig: symmetrization}. For the rest of this study we always use the harmonic time-step symmetrisation.

\subsection{Symmetrised free-fall and fly-by time-steps}

Next we provide the formulas for the time-step functions used in this study. In general the individual time-steps $\tau_\mathrm{i}$ assigned to the simulation particles must be shorter than the time-scale over which the orbits of the particles evolve (e.g. \citealt{Dehnen2011}). As our code is intended for collisional N-body simulations it is natural that the time-steps should be determined by the timescale of the close encounters the particles frequently experience. 

Following \cite{Pelupessy2012} we consider two simple time-step criteria based on two-body timescales: the free-fall timescale $t_\mathrm{ff}$ and the fly-by timescale $t_\mathrm{fb}$. We define the two-body free-fall time-step $\tau_\mathrm{ff}$ for particle $\mathcal{P}_\mathrm{i}$ as 
\begin{equation}\label{eq: dt-freefall}
\tau_\mathrm{ff,i} = \eta_\mathrm{ff}  \min_{\mathrm{j \neq i}} t_\mathrm{ff,ij}  = \eta_\mathrm{ff}  \min_{\mathrm{j \neq i}} \left( \frac{ \norm{ \vect{r}_\mathrm{ij} }^3 }{ G (m_\mathrm{i}+m_\mathrm{j}) }\right)^{1/2},
\end{equation}
in which the index $j$ runs over all other particles in the same level of the time-step hierarchy. Similarly, the fly-by time-step is defined as
\begin{equation}\label{eq: dt-flyby}
\tau_\mathrm{fb,i} = \eta_\mathrm{fb} \min_{\mathrm{j \neq i}} t_\mathrm{fb,ij}  = \eta_\mathrm{fb} \min_{\mathrm{j \neq i}} \frac{\norm{ \vect{r}_\mathrm{ij} }}{\norm{ \vect{v}_\mathrm{ij} }}.
\end{equation}
In the two equations the constants $\eta_\mathrm{ff}$ and $\eta_\mathrm{fb}$ are user-given integration accuracy parameters. In this study we always use $\eta = \eta_\mathrm{ff}=\eta_\mathrm{fb}$. With this definition the two time-step criteria agree on the time-step for a circular Keplerian binary i.e.
\begin{equation}\label{eq: eta-two-pi}
    \frac{\tau_\mathrm{ij}}{P} = \frac{\eta}{2 \pi},
\end{equation}
where $P$ is the orbital period of the binary. This expression provides a practical rule of thumb for estimating the time-step size compared to the orbital period as a function of the accuracy parameter $\eta$.

As our definition for a time-step is of the form $\tau_\mathrm{i} = \min_\mathrm{j}(t_\mathrm{ij})$, most importantly, containing a $\min$ function, we will first symmetrise the two-body timescales $t_\mathrm{ij}$ instead of the time-steps $\tau_\mathrm{i}$. The actual time-steps are finally obtained as a minimum of the symmetrised timescales as $\tau_\mathrm{i}^\mathrm{sym} = \min_\mathrm{j}(t_\mathrm{ij}^\mathrm{sym})$.

Finally we provide the expressions for the derivatives of the free-fall and fly-by timescales. The required derivatives (e.g. \citealt{Pelupessy2012}) are for the free-fall timescale
\begin{equation}\label{eq: dt-freefall-derivative}
\derfrac{t_\mathrm{ff,ij}}{t} = \frac{3}{2} \frac{ \vect{r}_\mathrm{ij} \cdot \vect{v}_\mathrm{ij} }{ \norm{\vect{r}_\mathrm{ij}}^2 } t_\mathrm{ff,ij}
\end{equation}
and the fly-by timescale
\begin{equation}\label{eq: dt-flyby-derivative}
\begin{split}
\derfrac{t_\mathrm{fb,ij}}{t} &= \left[ \frac{\vect{r}_\mathrm{ij}  \cdot \vect{v}_\mathrm{ij}}{ \norm{\vect{r}_\mathrm{ij}}^2   } -  \frac{\vect{v}_\mathrm{ij}  \cdot \vect{a}_\mathrm{ij}}{ \norm{\vect{v}_\mathrm{ij}}^2   }\right] t_\mathrm{fb}\\ &\approx \frac{\vect{r}_\mathrm{ij}  \cdot \vect{v}_\mathrm{ij}}{\norm{\vect{r}_\mathrm{ij}}^2} \left[ 1 + \frac{G(m_\mathrm{i}+m_\mathrm{j} )}{ \norm{\vect{v}_\mathrm{ij}}^2 \norm{\vect{r}_\mathrm{ij}}  }\right] t_\mathrm{fb,ij}.
\end{split}
\end{equation}
In the last expression we approximate that $\vect{a}_\mathrm{ij} \approx \vect{a}_\mathrm{ij}^{\mathrm{two-body}}$ i.e. that the local tidal field is weak. We note that this approximation may not be always valid leading to occasionally non-optimal symmetrised time-steps. The final symmetrised time-steps for the individual simulation particles can be obtained using the symmetrisation factor from Eq. \eqref{eq: time-step-factor} with the two-body timescales from Eq. \eqref{eq: dt-freefall} and Eq. \eqref{eq: dt-flyby} and their corresponding time derivatives.


\section{Numerical implementation of {\large\texttt{FROST}} in CUDA C for CPU-GPU clusters}\label{section: cuda}

\subsection{Why CUDA?}

Practically every direct-summation N-body code reaching particle numbers beyond a few times $10^5$ uses hardware acceleration in the form of GRAPE cards or GPUs (e.g. \citealt{Gaburov2009,Nitadori2012,Wang2015}). This is our approach for implementing our hierarchical fourth-order forward symplectic integrator HHS-FSI with symmetrised time-steps into the our new \frost{} code as well. We use the CUDA\footnote{NVIDIA Compute Unified Device Architecture, https://developer.nvidia.com/cuda-zone} C programming language. CUDA C allows for programming the bulk of the simulation code with a familiar C syntax for CPUs. The computationally intensive parts of the code such as $\bigO{N^2}$ force and time-step assignment loops are implemented as CUDA device kernels to be run on the GPU hardware.

Solving the N-body problem numerically using the direct summation approach is not an optimal task for GPUs considering the single instruction, multiple data (SIMD) architecture of the hardware. This is because the equations of motion of the individual simulation particles are coupled i.e. a single particle requires information about all the other particles residing in the GPU device memory. Despite this, GPU-accelerated direct summation codes show superior performance compared to correspondingly parallelised CPU codes. A common approach is to use the fast yet limited shared memory of the GPU device (e.g. \citealt{Nguyen2007}).


\subsection{Implementation of CUDA kernels for all-pairs \texorpdfstring{$\bigO{N^2}$ o} operations}

Our N-body code contains three distinct all-pairs $\bigO{N^2}$ operations over simulation particles: the calculation of Newtonian accelerations, the gradient accelerations and the time-step assignment. If the particle number $N$ exceeds a few thousand particles these operations are performed on GPUs. Since the three all-pairs operations are very similar in their implementations we present here only the case of the Newtonian particle accelerations.

\begin{algorithm}\label{alg: kernels}
  \SetKwFunction{proc}{particle\_particle\_acc}
  \SetKwProg{myproc}{CUDA kernel}{}{}
  \myproc{\proc{$\vect{r}_\mathrm{i}$,$\vect{r}_\mathrm{j}$,$m_\mathrm{j}$,$\vect{a}_\mathrm{i}$}}
    {
    evaluate $\vect{a}$ from $\vect{r}_\mathrm{j}-\vect{r}_\mathrm{i}$ and $m_\mathrm{j}$ using Eq. \eqref{eq: acc-newton}\;
    $\vect{a}_\mathrm{i} \gets \vect{a}_\mathrm{i} + \vect{a}$\;
    \KwRet $\vect{a}_\mathrm{i}$ \;
    }
  \SetKwFunction{procc}{particle\_tile\_acc}
  \SetKwProg{myprocc}{CUDA kernel}{}{}
  \myprocc{\procc{$\vect{r}_\mathrm{i}$, $\vect{a}_\mathrm{i}$}}
    {
    \While{$j=1:\mathrm{num\_threads}$}
    {
    $\vect{r}_\mathrm{j} \gets \mathrm{shared\_memory\_r} \left[ j \right]$\;
    $m_\mathrm{j} \gets \mathrm{shared\_memory\_m}\left[j\right]$\;
    $\vect{a}_\mathrm{i} \gets$ \proc{$\vect{r}_\mathrm{i}$,$\vect{r}_\mathrm{j}$,$m_\mathrm{j}$,$\vect{a}_\mathrm{i}$}
    }
    \KwRet $\vect{a}_\mathrm{i}$\;
    }
    
  \SetKwFunction{proccc}{particle\_acc}
  \SetKwProg{myproccc}{CUDA kernel}{}{}
  \myproccc{\proccc{$\{\vect{r}_\mathrm{j}\}$, $\{\vect{a}_\mathrm{j}\}$}}
    {
    $N\gets$ length( $\{ \vect{a}_\mathrm{j} \}$ )\;
    $\vect{a}_\mathrm{thread} \gets 0$\;
    $\mathrm{tid} \gets \mathrm{this\_block\_id}\times\mathrm{num\_threads} + \mathrm{this\_thread\_id}$\;
    $\vect{r}_\mathrm{thread} \gets \vect{r}_\mathrm{tid}$\;
    
    \While{$k=1:\mathrm{N}/\mathrm{num\_threads}$}{
    \hspace{-0.15cm}$\mathrm{idx} = k \times \mathrm{num\_threads}+\mathrm{this\_thread\_id}$\;
    \hspace{-0.15cm}$\mathrm{shared\_memory\_r}\left[\mathrm{this\_thread\_id}\right] \gets \vect{r}_\mathrm{idx}$\;
    \hspace{-0.15cm}$\mathrm{shared\_memory\_m}\left[\mathrm{this\_thread\_id}\right] \gets m_\mathrm{idx}$\;
    \hspace{-0.15cm}$\vect{a}_\mathrm{thread} \gets$ \procc{$\vect{r}_\mathrm{thread}$, $\vect{a}_\mathrm{thread}$ }\;
    }
    \If{ $\mathrm{tid}<N$ }{
        $\vect{a}_\mathrm{tid} \gets \vect{a}_\mathrm{thread}$\;
    }
    } 
  \SetKwFunction{procccc}{compute\_acc\_newton}
  \SetKwProg{myprocccc}{C function}{}{}
  \myprocccc{\procccc{ $\{\vect{r}_\mathrm{i}\}$, $\{m_\mathrm{i}\}$, $\{\vect{a}_\mathrm{i}\}$ }{}}
    {
    \hspace{-0.05cm}$N\gets$ length( $\{ \vect{a}_\mathrm{i} \} )$\;
    \hspace{-0.05cm}copy \{$\vect{r}_\mathrm{i}$, $m_\mathrm{i}$\} from CPU to GPU memory\;
    \hspace{-0.05cm}$p \gets $ num\_threads, e.g. $32$--$128$\;
    \hspace{-0.05cm}$q \gets (p+1)/p\times N$\;
     \hspace{-0.05cm}launch kernel \procc{ $\{\vect{r}_\mathrm{i}\}$,$\{m_\mathrm{i}\}$,$\{\vect{a}_\mathrm{i}\}$ } with $q$ blocks\_per\_grid \& $p$ threads\_per\_block\;
    \hspace{-0.05cm}copy \{$\vect{a}_\mathrm{i}$\} from GPU to CPU memory\;
    \hspace{-0.05cm}\KwRet $\{\vect{a}_\mathrm{i}\}$\;
    }
  \caption{A pseudocode for computing all-pair Newtonian accelerations for $N$ particles with CUDA C. The acceleration calculation on GPUs begins when the C function \texttt{compute\_acc\_newton()} calls the CUDA kernel \texttt{particle\_acc()} which then in turn uses the other two particle acceleration kernels \texttt{particle\_tile\_acc()} and \texttt{particle\_particle\_acc()} as explained in the text. The computation of time-steps and the gradient accelerations are performed in an analogous manner.}
\end{algorithm}

\begin{figure}
\includegraphics[width=\linewidth]{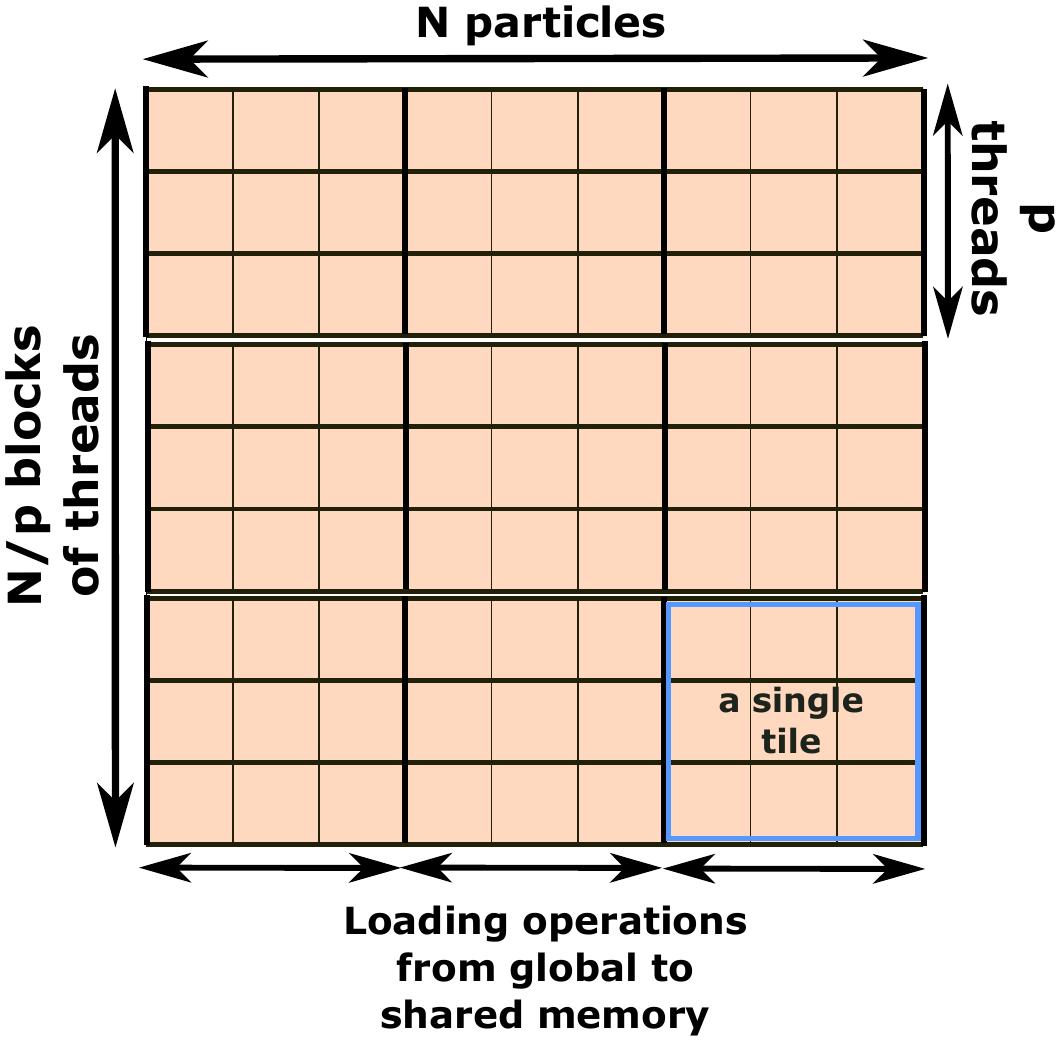}
\caption{The GPU force computation for $9$ particles. The figure illustrates a grid of thread blocks consisting of tiles loaded into the shared memory of the GPU device. The required 81 force computations are sorted into 9 tiles with 9 interactions each executed on three GPU threads. The figure is outlined following the diagrams of \protect\cite{Nguyen2007}. See also the illustrations of \protect\cite{Gaburov2009}.}
\label{fig: grid}
\end{figure}

The calculation of the Newtonian accelerations on GPUs is presented in pseudocode in Algorithm \eqref{alg: kernels}. The algorithm contains three CUDA kernels for the actual calculations and a single C function for launching the kernels. The C function is also responsible for copying the data between the CPU host and the GPU device memories.

We launch the global CUDA kernel \texttt{particle\_acc()} using typically $p=32$ threads per block and $q=\floor{(p+1)/p} N$ blocks per grid in which $N$ is the number of particles. A single thread computes the acceleration for a single particle. These particles are referred to as i-particles \citep{Gaburov2009}. In order to speed up the memory access in the GPU code we use the fast (and limited) shared memory of the GPU device. A basic unit for computing partial accelerations for $p$ particles is a tile of j-particles loaded into the shared memory of the device. All threads in the same thread block can access the same shared memory. See Fig. \ref{fig: grid} for a schematic illustration of tiles, threads and blocks in an all-pairs operation. The threads proceed calling the following acceleration CUDA kernels and loading subsequent tiles into the shared memory until all $N$ j-particles have been processed for each i-particle. 
\begin{itemize}
    \item Kernel \texttt{particle\_tile\_acc($\vect{r}_\mathrm{i},\vect{a}_\mathrm{i}$)}\\ A single tile is used to compute accelerations for $p$ i-particles from $p$ j-particles in the shared memory of the thread block. The kernel essentially loops through the shared memory and loads new j-particles for the particle-particle acceleration calculation kernel below.
    \item Kernel \texttt{particle\_particle\_acc($\vect{r}_\mathrm{i}, \vect{r}_\mathrm{j}, m_\mathrm{j} ,\vect{a}_\mathrm{i}$)}\\ The innermost CUDA kernel calculating the Newtonian particle accelerations. The kernel calculates the acceleration of the i-particle due to the j-particle and adds it in the total acceleration of the i-particle.
\end{itemize}
We have also implemented somewhat more complex multi-thread algorithm \citep{Nguyen2007} which speeds up the acceleration calculation by a few tens of percents especially when the particle number and thus the GPU occupancy is low. The essence of the multi-threaded algorithm is that multiple threads participate in the acceleration computation of a single i-particle. After the accelerations have been obtained the threads in the same block use the shared device memory to sum the total acceleration of the i-particle.

For running the code on multiple GPUs assigned to different computing cluster nodes, inter-node communication is necessary. We employ the widely-used MPI\footnote{Message Passing Interface, https://www.mpi-forum.org/} standard for hybrid MPI-CUDA parallelisation of the all-pairs operations. Throughout this study we use one MPI task per one GPU device employing the common scatter-compute-gather communication scheme for parallelising computationally expensive parts of the code. MPI is also used for CPU loop parallelisation if the particle number is too low and GPUs cannot be used efficiently. Finally, serial CPU code is used when $N\lesssim$ a few hundred particles. 

\subsection{Implementation of the FSI algorithm}

The plain forward symplectic integrator algorithm (without hierarchical Hamiltonian splitting) is implemented as the FSI function of our code. The algorithm is presented in pseudocode in Algorithm \eqref{alg: fsi}.

\begin{algorithm}\label{alg: fsi}
\SetKwFunction{proc}{fsi}
\SetKwProg{myproc}{C function}{}{}
\myproc{\proc{ $\{m_\mathrm{i}\}$, $\{\vect{r}_\mathrm{i}\}$, $\{\vect{v}_\mathrm{i}\}$, $\Delta t$}}
{
\If{$\{\{m_\mathrm{i}\}$,$\{\vect{r}_\mathrm{i}\}$,$\{\vect{v}_\mathrm{i}\}\}$ $\neq$ $\varnothing$
}{
$\{\vect{a}_\mathrm{i}\} \gets$\texttt{compute\_acc\_newton($\{\vect{r}_\mathrm{i}\}$,$\{m_\mathrm{i}\}$)}\;
\texttt{kick}($\{\vect{v}_\mathrm{i}\}$,$\{\vect{a}_\mathrm{i}\}$, $\sfrac{1}{6}\,\Delta t$)\;
\vspace{0.04cm}
\texttt{drift}($\{\vect{r}_\mathrm{i}\}$,$\{\vect{v}_\mathrm{i}$\}, $\sfrac{1}{2}\,\Delta t$)\;
\vspace{0.04cm}
$\{\vect{\tilde{a}}_\mathrm{i}\} \gets$\texttt{compute\_acc\_gradient($\{\vect{r}_\mathrm{i}\}$,$\{m_\mathrm{i}\}$)}\;
\texttt{kick}($\{\vect{v}_\mathrm{i}\}$,$\{\vect{\tilde{a}}_\mathrm{i}\}$, $\sfrac{2}{3}\,\Delta t$)\;
\vspace{0.04cm}
\texttt{drift}($\{\vect{r}_\mathrm{i}\}$,$\{\vect{v}_\mathrm{i}$\}, $\sfrac{1}{2}\,\Delta t$)\;
\vspace{0.04cm}
$\{\vect{a}_\mathrm{i}\} \gets$\texttt{compute\_acc\_newton($\{\vect{r}_\mathrm{i}\}$,$\{m_\mathrm{i}\}$)}\;
\texttt{kick}($\{\vect{v}_\mathrm{i}\}$,$\{\vect{a}_\mathrm{i}\}$, $\sfrac{1}{6}\,\Delta t$)\;
}
\KwRet $\{\{m_\mathrm{i}\},\{\vect{r}_\mathrm{i}\}$,$\{\vect{v}_\mathrm{i}$\}\}\;
}
\caption{The pseudocode implementation of the FSI integrator of Eq. \eqref{eq: fsi}.}
\end{algorithm}
The integrator evolves a given N-body system according to the time evolution operator of Eq. \eqref{eq: fsi}. FSI is the only function of the \frost{} code which actually propagates the particle positions forward in time. The function contains the following standard integration operations: \texttt{drift()} and \texttt{kick()} which are discussed in detail below.
\begin{itemize}
    \item Function \texttt{drift}(\{$\vect{r}_\mathrm{i}$\}, \{$\vect{v}_\mathrm{i}$\}, $\Delta t$)\\ Propagates the individual particles from the position $\{\vect{r}_\mathrm{i}\}$ into \{$\vect{r}_\mathrm{i} + \Delta t\, \vect{v}_\mathrm{i}$\}.
    \item Function \texttt{kick}(\{$\vect{v}_\mathrm{i}\}$, \{$\vect{a}_\mathrm{i}$\}, $\Delta t$\})\\ Updates the individual particle velocities from $\{\vect{v}_\mathrm{i}\}$ to \{$\vect{v}_\mathrm{i} + \Delta t\, \vect{a}_\mathrm{i}$\} using given accelerations \{$\vect{a}_\mathrm{i}$\} (Newtonian or gradient) computed using Algorithm \eqref{alg: kernels} or its gradient counterpart from Eq. \eqref{eq: acc-gradient}. A single all-pairs $\bigO{N^2}$ operation is required and GPU acceleration is used to speed up the calculation.
\end{itemize}

\subsection{Implementation of the HHS-FSI algorithm}

The hierarchical Hamiltonian splitting approach of our integrator HHS-FSI manifests itself in the recursive nature of the \texttt{hhs\_fsi()} function. Most importantly, the function performs the splitting of the simulation particles into two sets using a pivot time-step. The set of slow particles is integrated by calling the previously presented \texttt{fsi()} function in Algorithm \eqref{alg: fsi}. The set of fast particles is inserted again into \texttt{hhs\_fsi()} for further hierarchical integration. The function \texttt{hhs\_fsi()} is described in pseudocode in Algorithm \eqref{alg: hhs-fsi}.

\begin{algorithm}\label{alg: hhs-fsi}
\SetKwFunction{proctwo}{hhs\_fsi}
\SetKwProg{myproctwo}{C function}{}{}
\myproctwo{\proctwo{ $\{m_\mathrm{i}\}$,$\{\vect{r}_\mathrm{i}\}$,$\{\vect{v}_\mathrm{i}\}$, $\tau_\mathrm{pivot}$  }}
{
\If{ $\{\{m_\mathrm{i}\}$,$\{\vect{r}_\mathrm{i}\}$,$\{\vect{v}_\mathrm{i}\}\}$ $\neq$ $\varnothing$ }{
\BlankLine
$\{\tau_\mathrm{i}\} \gets$ \texttt{assign\_timesteps}($\{m_\mathrm{i}\}$,$\{\vect{r}_\mathrm{i}\}$,$\{\vect{v}_\mathrm{i}\}$)\;
\vspace{-0.4cm} \begin{equation*}\hspace{-0.3cm}
\left.\begin{aligned}
&\mathcal{S} \equiv \{\{m_\mathrm{j}\},\{\vect{r}_\mathrm{j}\},\{\vect{v}_\mathrm{j}\}\}_\mathrm{S}\\
&\mathcal{F} \equiv\{\{m_\mathrm{k}\},\{\vect{r}_\mathrm{k}\},\{\vect{v}_\mathrm{k}\}\}_\mathrm{F}
\end{aligned}
 \right\}
 \gets \text{\texttt{partition}}(\tau_\mathrm{pivot}, \{\tau_\mathrm{i}\})\text{;}
\end{equation*}
$\{\{\vect{a}_\mathrm{j}\},\{\vect{a}_\mathrm{k}\}\} \gets$ \texttt{acc\_sf\_newton($\mathcal{S},\mathcal{F}\,$)}\;
\vspace{0.04cm}
\texttt{kick\_sf}(\{$\{\vect{v}_\mathrm{j}\},\{\vect{a}_\mathrm{j}\}\}_{\mathrm{S}}$,$\{\{\vect{v}_\mathrm{k}\},\{\vect{a}_\mathrm{k}\}\}_{\mathrm{F}}$,$\sfrac{1}{6}\,\tau_\mathrm{pivot}$)\;
\vspace{0.15cm}
\texttt{fsi}($\mathcal{S}$,$\sfrac{1}{2}\,\tau_\mathrm{pivot}$)\;
\vspace{0.04cm}
\texttt{hhs\_fsi}($\mathcal{F}$,$\sfrac{1}{2}\,\tau_\mathrm{pivot}$)\;
\vspace{0.15cm}
$\{\{\vect{\tilde{a}}_\mathrm{j}\},\{\vect{\tilde{a}}_\mathrm{k}\}\} \gets$ \texttt{acc\_sf\_gradient($\mathcal{S},\mathcal{F}\,$)}\;
\texttt{kick\_sf}(\{$\{\vect{v}_\mathrm{j}\},\{\vect{\tilde{a}}_\mathrm{j}\}\}_{\mathrm{S}}$,$\{\{\vect{v}_\mathrm{k}\},\{\vect{\tilde{a}}_\mathrm{k}\}\}_{\mathrm{F}}$,$\sfrac{2}{3}\,\tau_\mathrm{pivot}$)\;
\vspace{0.15cm}
\texttt{hhs\_fsi}($\mathcal{F}$,$\sfrac{1}{2}\,\tau_\mathrm{pivot}$)\;
\vspace{0.04cm}
\texttt{fsi}($\mathcal{S}$,$\sfrac{1}{2}\,\tau_\mathrm{pivot}$)\;
\vspace{0.15cm}
$\{\{\vect{a}_\mathrm{j}\},\{\vect{a}_\mathrm{k}\}\} \gets$ \texttt{acc\_sf\_newton($\mathcal{S},\mathcal{F}\,$)}\;
\texttt{kick\_sf}(\{$\{\vect{v}_\mathrm{j}\},\{\vect{a}_\mathrm{j}\}\}_{\mathrm{S}}$,$\{\{\vect{v}_\mathrm{k}\},\{\vect{a}_\mathrm{k}\}\}_{\mathrm{F}}$,$\sfrac{1}{6}\,\tau_\mathrm{pivot}$)\;
\vspace{0.2cm}
$\{\{m_\mathrm{i}\}$,$\{\vect{r}_\mathrm{i}\}$,$\{\vect{v}_\mathrm{i}\}\} \gets \mathcal{S} \cup \mathcal{F}$\;
}
\KwRet $\{\{m_\mathrm{i}\}$,$\{\vect{r}_\mathrm{i}\}$,$\{\vect{v}_\mathrm{i}\}\}$\;
}
\caption{The pseudocode implementation of the HHS-FSI integrator of Eq. \eqref{eq: time-evolution-operator-hhs-fsi}. The sets $\mathcal{S}$ and $\mathcal{F}$ are a shorthand notation for the sets of slow and fast particles, respectively. Note the recursive nature of the algorithm which manifests the hierarchical nature of the HHS-FSI integrator.}
\end{algorithm}

The functions \texttt{hhs\_fsi()} calls during its execution are detailed in the list below.
\begin{itemize}
    \item Function \texttt{assign\_timesteps}($\{\{m_\mathrm{i}\},\{\vect{r}_\mathrm{i}\},\{\vect{v}_\mathrm{i}\}\}$)\\
    The time-step assignment function computes and symmetrises the free-fall and fly-by time-steps and chooses the shortest step for each particle using \Cref{eq: dt-freefall,eq: dt-flyby,eq: time-step-factor,eq: dt-freefall-derivative,eq: dt-flyby-derivative}.
    
    \item Function \texttt{partition}($\tau_\mathrm{pivot}$, $\{\tau_\mathrm{i}\}$)\\
    This function partitions the set of particles gives as its input into two particle sets: slow and fast particles. A particle belongs to the set of slow particles if $\tau_\mathrm{i}\geq\tau_\mathrm{pivot}$ i.e. its time-step is longer than the given pivot step. If not, the particle belongs to the set of fast particles. The union of the two particle subsets is always equivalent to the original set of particles. Note that either one (but not both) of the slow and fast particle sets may be an empty set. 
    
    \item Function \texttt{kick\_sf}($\{\{\vect{v}_\mathrm{j}\}$,$\{\vect{a}_\mathrm{j}\}\}_{\mathrm{S}}$,\,$\{\{\vect{v}_\mathrm{k}\}$,$\{\vect{a}_\mathrm{k}\}\}_{\mathrm{F}}$,\,$\tau_\mathrm{pivot}$)\\
    The function performs the pairwise kicks between the particles on different slow and fast levels in the integration hierarchy. 

    \item Function \texttt{acc\_sf\_newton}(\{$\{m_\mathrm{j}\},\{\vect{{r}}_\mathrm{j}\}\}_{\mathrm{S}}$,$\{\{m_\mathrm{k}\},\{\vect{{r}}_\mathrm{k}\}\}_{\mathrm{F}}$);\\
    The Newtonian inter-level accelerations for the kicks are computed using Eq. \eqref{eq: force-newton-inter}. Note that particles on the same hierarchy level do not interact within the function. GPU acceleration is used to speed up the calculation as explained before.
    
    \item Function
    \texttt{acc\_sf\_gradient}(\{$\{m_\mathrm{j}\},\{\vect{r}_\mathrm{j}\}\}_{\mathrm{S}}$,$\{\{m_\mathrm{k}\},\{\vect{r}_\mathrm{k}\}\}_{\mathrm{F}})$;\\
    Analogous to the function above, this function carries out the computation of the pairwise gradient accelerations between particles on slow and fast levels of the time-step hierarchy. The inter-level gradient accelerations are calculated from Eq. \eqref{eq: force-gradient-inter-1} and Eq. \eqref{eq: force-gradient-inter-2}. GPUs are employed for the two expensive pairwise acceleration computations.
\end{itemize}

\subsection{Basic structure of the {\large\texttt{FROST}} code}

The main function level of the \frost{} code contains the standard initialisation of a MPI-parallelised CUDA C program, the memory management functions, the input and output (IO) and the main simulation loop of the code. The main loop is responsible for running the simulation itself from the start time $t_\mathrm{start}$ to stop time $t_\mathrm{stop}$ in intervals of $\Delta t$ which is the first (and longest) pivot step $\tau_\mathrm{pivot}$. The integration interval is also the first pivot time-step given to the integrator function \texttt{hhs\_fsi()} and corresponds to the maximum time-step in block time-step codes. The pseudocode of the main function of the \frost{} code is provided in Algorithm \eqref{alg: frost}.
\begin{algorithm}\label{alg: frost}
\SetKwFunction{proctwo}{\texttt{FROST}}
\SetKwProg{myproctwo}{main C function}{}{}
\myproctwo{\proctwo{$\mathrm{parameter\_file}$}}
{
\hspace{-2mm}\texttt{initialise\_cuda\_and\_mpi()}\;
\hspace{-2.5mm}\{ic\_file, $t_\mathrm{start}, t_\mathrm{end}$, $\Delta t$\} $\gets$  \texttt{read\_input}(parameter\_file)\;
\hspace{-2mm}\texttt{allocate\_memory()}\;
\hspace{-2mm}\{$\{m_\mathrm{i}\}$,$\{\vect{r}_\mathrm{i}\}$,$\{\vect{v}_\mathrm{i}\}$\} $\gets$ \texttt{read\_ic\_file}(ic\_file)\;
\BlankLine
\hspace{-2mm}$t \gets t_\mathrm{start}$\;
\hspace{-2mm}\While{ $t<t_\mathrm{end}$ }{
\hspace{-2mm}\texttt{hhs\_fsi}( \{$\{m_\mathrm{i}\}$,$\{\vect{r}_\mathrm{i}\}$,$\{\vect{v}_\mathrm{i}\}$\}, $\Delta t$ )\;
$t \gets t+\Delta t$\;
\hspace{-2mm}\texttt{on\_the\_fly\_analysis()}\;
\hspace{-2mm}\texttt{write\_snapshot\_file\_if\_desired()}\;
}
\hspace{-2mm}\texttt{free\_memory()}\;
\hspace{-2mm}\texttt{finalise\_mpi()}
}\caption{The \frost{} code main integration loop.}
\end{algorithm}

The functions on the main loop level of the \frost{} code are described in detail below.
\begin{itemize}
    \item \texttt{initialise\_cuda\_and\_mpi()}, \texttt{finalise\_mpi()}\\
    The standard initialisation and termination of the MPI library access. Each MPI is bind to a single GPU in this function as well. 
    \item \texttt{allocate\_memory()}, \texttt{free\_memory()}\\
    The dynamic memory allocation (and freeing) for arrays of variables both in the CPU host memory and the GPU device memory. In our code we use the CUDA memory allocation also for allocating the host memory. 
    \item \texttt{read\_input}(parameter\_file), \texttt{read\_ic\_file}(ic\_file)\\
    Functions for reading the user-given parameters and the initial conditions for the simulation.
    \item \texttt{write\_snapshot\_file\_if\_desired()}\\
    If user-given amount of simulation time has elapsed since writing the previous snapshot file this function writes a new snapshot file. The format of the snapshot file is identical to the format of the ic\_file so snapshots can be used to restart and continue a simulation. 
    \item \texttt{on\_the\_fly\_analysis()}\\
    Performs simulation analysis which requires such a high time resolution that the analysis from written snapshots afterwards would consume an impractically large amount of disk space. Typical examples are monitoring the conservation of energy, momentum and angular momentum i.e. the numerical accuracy of the simulation, or saving the global physical properties of the simulated N-body system, e.g. Lagrangian radii, virial parameter or statistics of bound binaries.
\end{itemize}


\section{Integrator performance}\label{section: 6}
\subsection{Few-body simulations}
\subsubsection{Keplerian binaries}

\begin{figure}
\includegraphics[width=\linewidth]{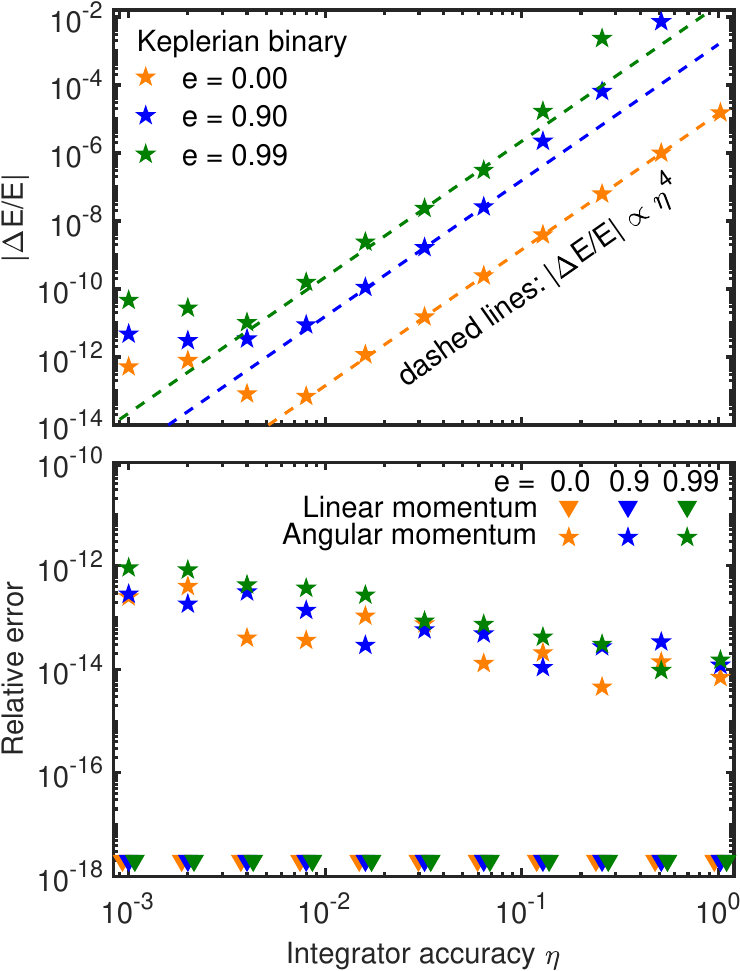}
\caption{Error analysis of the Keplerian binary experiments. The colors (orange, blue, green) correspond to binary orbits of $10 M_{\odot}$ stars with 10 AU semi-major axis and eccentricities $e=0.0$, $e=0.9$ and $e=0.99$, respectively. Top panel: the relative energy error $|\Delta E / E|$ as a function of the integrator accuracy parameter $\eta$. The energy error follows closely the expected $|\Delta E / E| \propto \eta^4$ behaviour. At very high accuracy ( small $\eta$) round-off errors become dominant. The relative energy error is larger than expected from the scaling for highly eccentric binaries at $\eta \gtrsim 0.1$ as the pericenter passages are not properly resolved. Bottom panel: the relative (linear) momentum $|\Delta P / P|$ and angular momentum error $|\Delta L / L|$. The momentum is exactly conserved down to machine precision as $\Delta P = 0$. The angular momentum error is governed by accumulating floating-point round-off errors and is similar for all three binary eccentricities with a maximum value of $|\Delta L / L| \sim 10^{-12}$.}
\label{fig: binary}
\end{figure}
We setup Keplerian point-mass binaries with binary component masses of $m_\mathrm{1}=m_\mathrm{2}=10 \, M_\mathrm{\odot}$ and semi-major axis of $10 \, \mathrm{AU}$. Three different orbital eccentricities are used: $e=0.00$, $e=0.90$ and $e=0.99$. In order to investigate the numerical performance of \frost{} we integrate the two-body systems for $1000$ orbital periods and examine the conservation of total energy $E$, total linear momentum $P=\norm{\vect{P}}$ and total angular momentum $L = \norm{\vect{L}}$. In total $11$ different integration accuracy parameters $\eta$ from the interval $0.001 \leq \eta \leq 1.024$ are used. Considering Eq. \eqref{eq: eta-two-pi} the maximum time-step corresponds to approximately one sixth of the orbital period of the binary. Due to the mutual nature of the time-steps both particles always share the same level in the time-step hierarchy. Thus, two-body experiments only assess the performance of the CPU implementation of FSI in \frost, not the full HHS-FSI integrator.

The results of the Keplerian binary runs are gathered in Fig. \ref{fig: binary}. The top panel of the figure shows the relative energy error $|\Delta E / E| \equiv |(E_\mathrm{t=1000P}-E_\mathrm{t=0})/E_\mathrm{t=0}|$ as function of time for all the $33$ two-body runs. Beginning from the circular ($e=0.0$) runs we see that the relative energy error closely follows the relation $|\Delta E / E| \propto \eta^4$ between the accuracy parameter values of $0.008 \leq \eta \leq 1.024$. This fact confirms the order of the FSI in \frost{} as $|\Delta E / E| \propto \eta^4$ is the expected behaviour for a fourth-order integrator \citep{Dehnen2017a}. At $\eta \sim 0.008$ the relative energy error is $|\Delta E / E \sim 10^{-13}|$. With smaller values of the accuracy parameter i.e. $\eta<0.008$ the floating-point round-off errors begin to dominate and the $|\Delta E / E|$ begins to increase again. Thus, there is an optimal finite value for $\eta$ for reaching minimum energy error depending on the system studied.

The runs with eccentric binaries $e=0.9$ and $e=0.99$ behave qualitatively similarly as the $e=0.0$ case when $\eta \lesssim 0.1$. At small values of $\eta$ the floating-point round-off error dominates, the minimum relative energy error $|\Delta E / E \sim 10^{-12}$ -- $10^{-10}$ is obtained at $\eta \sim 0.008$ after which $|\Delta E / E| \propto \eta^4$. The behaviour of the relative energy error deviates from the fourth-order scaling at $\eta \gtrsim 0.1$ i.e. the error is larger than what is expected from the fourth-order scaling. The reason for the increased error is that the time-steps become too large for properly resolving the rapid close pericenter passages of the bodies in eccentric binaries.

The bottom panel of in Fig. \ref{fig: binary} shows the relative errors of linear momentum $|\Delta P / P|$ and angular momentum $|\Delta L / L|$ for the three different orbital eccentricities and $11$ integration accuracy parameters $\eta$. The relative errors of $P$ and $L$ are defined analogously to the relative energy error above. In the binary runs we observe an exact conservation withing numerical precision ($\Delta P = 0$) which confirms the momentum conservation of our implementation of the FSI. The relative angular momentum error is $|\Delta L / L|$ is governed by the floating-point errors, increasing towards smaller values of $\eta$ i.e. larger number of taken steps and floating-point operations. However, the maximum relative angular momentum error is still very small, $|\Delta L / L| \lesssim 10^{-12}$. There are no apparent differences in angular momentum conservation between the three binary eccentricities.

\subsubsection{Systems with a dominant central body}

\begin{figure}
\includegraphics[width=\linewidth]{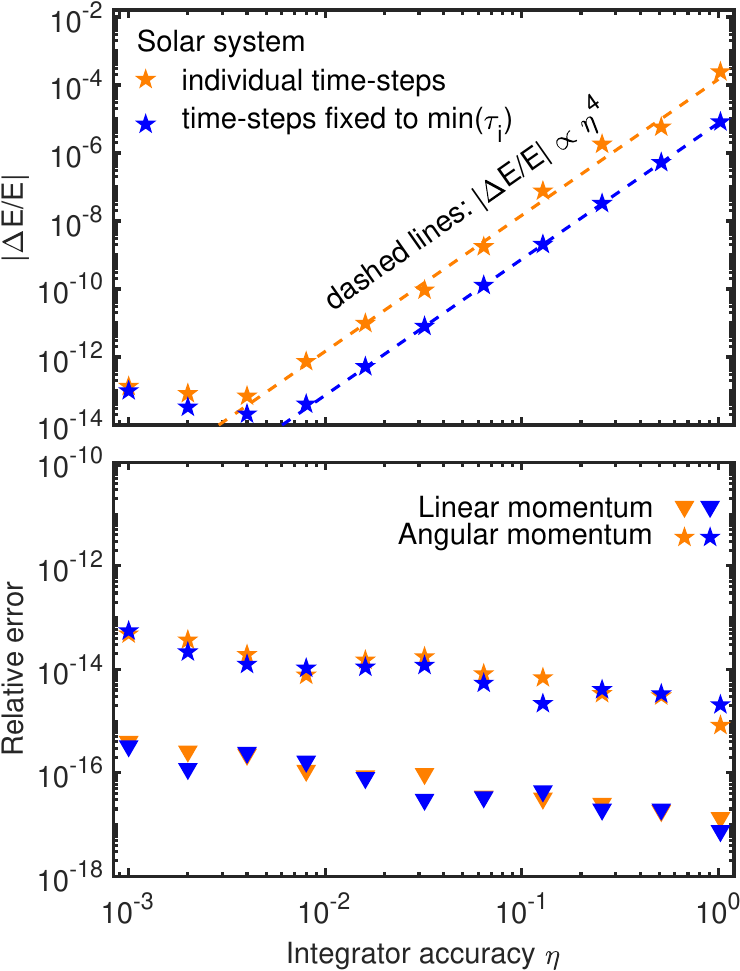}
\caption{Error analysis of the five-body Solar system experiments. See text for the simulation setup. In contrast to the Keplerian binary test Fig. \ref{fig: binary} with fixed time-steps, the Solar system setup tests the hierarchical HHS-FSI integration with individual particle time-steps (here in orange). The Solar systems runs with fixed time-steps are show in blue. Top panel: the relative energy error $|\Delta E / E|$ as a function of the integrator accuracy parameter $\eta$. When $\eta \gtrsim 0.004$ the energy error follows the power-law $|\Delta E / E| \propto  \eta^4$ as expected from a fourth-order integrator. With lower values of $\eta$ the floating-point round-off error again dominates. Bottom panel: the relative linear momentum and angular momentum conservation is determined by the round-off error as the error increases towards small values of $\eta$. Note that overall the errors are very small (for energy when $\eta\lesssim0.1$), for example at $\eta=0.004$ we have $|\Delta E / E| \sim 10^{-13}$, $|\Delta L / L| \sim 10^{-14}$ and $|\Delta P / P| \sim 10^{-16}$.}
\label{fig: solarsystem}
\end{figure}
We perform another series of few-body experiments to evaluate the accuracy and confirm the order of the HHS-FSI integrator of \frost. A good test setup is a solar system consisting of a dominant central mass (star) and a collection of orbiting low-mass bodies (planets). If the semi-major axes of the planets w.r.t. the star are different enough the planets will end up in different levels on the time-step hierarchy with the star sharing the fastest level with the innermost planet. Thus, this setup also tests the inter-level interactions unlike the two-body experiments above.

We choose our star, the Sun, and the four giant planets of the Solar system as the initial conditions of the five-body experiments. See \cite{Dehnen2017a} and Appendix \ref{appendix: planets} for the exact initial state of the system. We run the Solar system initial conditions for $1000$ years with the integrator accuracy parameters $\eta$ in the range $0.001 \leq \eta \leq 1.024$, just as in the Keplerian binary experiments. In addition to the tests with the HHS-FSI integrator we perform another set of runs in which all the five particles are forced to the the fastest hierarchy level i.e. the minimum time-step. This procedure results in five-body FSI integration as the hierarchical nature of the integration is removed.

The final results of the five-body Solar system experiments are displayed in Fig. \ref{fig: solarsystem}. The results are qualitatively similar to the case of circular binaries in the previous section as the osculating orbital eccentricities of the giant planets in our Solar system are low\footnote{The JPL Solar System homepage https://ssd.jpl.nasa.gov/,\\orbital elements from https://ssd.jpl.nasa.gov/txt/p\_elem\_t1.txt.}, typically $e\lesssim 0.01$. The relative energy error  (top panel) again follows the expected fourth-order relation $|\Delta E / E| \propto \eta^4$ when $\eta\gtrsim0.004$, confirming that our implementation of the novel HHS-FSI is indeed a fourth-order integrator. Below $\eta=0.004$ the round-off error again governs the error behaviour of the runs. In the FSI runs with all particles set to the fastest hierarchy level the relative energy errors are approximately an order of magnitude smaller than in the HHS-FSI simulations when round-off error does not dominate. However, the cost of not using the hierarchical integration is the increased running time due to which equal time-step runs become impractical when the particle number is large.

In the bottom panel of Fig. \ref{fig: solarsystem} we see that the round-off error again dictates the behaviour of the relative angular momentum error $|\Delta L / L|$ with less error towards higher values of $\eta$. The maximum relative angular momentum error is still small, less than $10^{-13}$. However, now the linear momentum is not exactly conserved anymore i.e. $|\Delta P|>0$ and behaves similarly as $|\Delta L / L|$ due to floating-point round-off error as there are multiple acceleration vectors to sum for $N>2$ bodies. The linear momentum error is always extremely small, $|\Delta P|<10^{-15}$. The results of the fixed minimum time-steps simulation set do not differ from the HHS-FSI runs for linear and angular momentum.

\subsection{Million-body simulations}

\subsubsection{Conservation of energy, momentum and angular momentum}\label{section: million-conservation}

\begin{figure*}
\includegraphics[width=\textwidth]{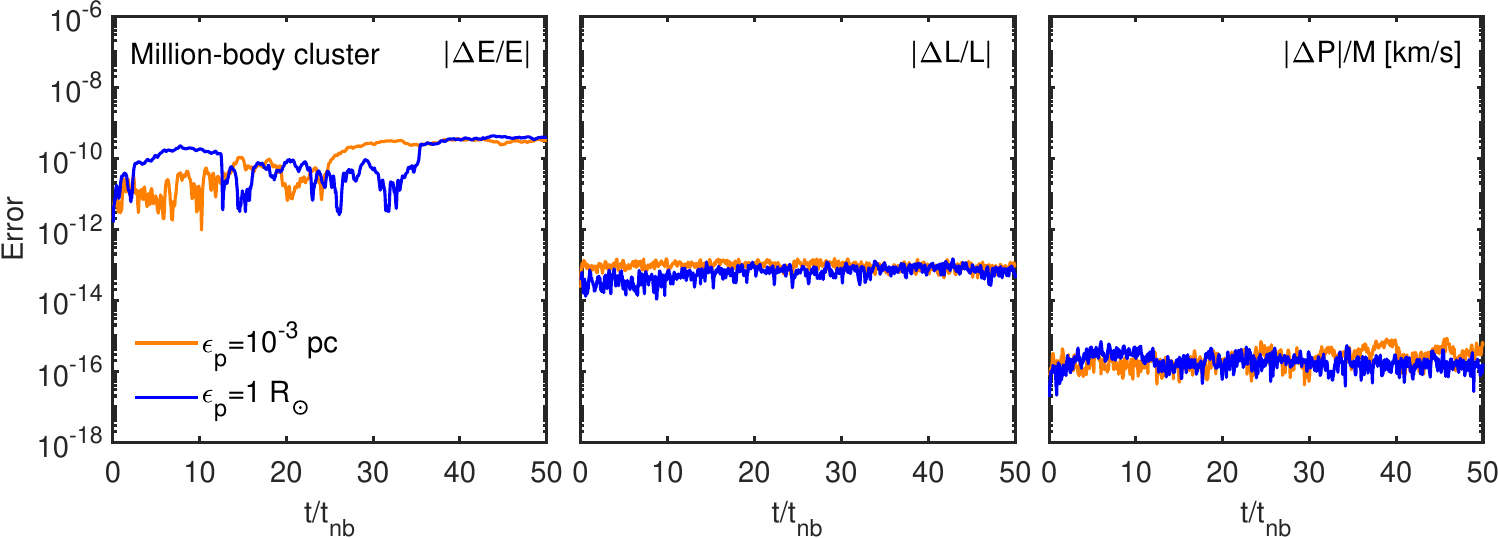}
\caption{Error analysis for a star cluster realised with $1\times 10^6$ stars. See the text for the initial setup. From left to right we show the relative energy error $|\Delta E / E|$, the angular momentum error $|\Delta L / L|$ and the linear momentum error as $|\Delta P|/M$ using $\eta=0.2$ with gravitational softening lengths of $\epsilon_\mathrm{P} = 10^{-3}$ pc (orange line) and $\epsilon_\mathrm{P} = 1\,R_\mathrm{\odot} \sim 2.3\times10^{-8}$ pc (in blue). There are no qualitative differences between results with the two different gravitational softening lengths. The relative energy error $|\Delta E / E|$ (left panel) fluctuates initially. After $50$ N-body time units ($\sim 12$ Myr) the relative energy error is only $|\Delta E / E| \sim 2 \times 10^{-10}$. The relative angular momentum error (middle panel) remains approximately constant around $|\Delta L / L| \sim 10^{-13}$. The absolute error of the linear momentum (right panel) corresponds to the center-of-mass velocity $|\Delta P|/M \sim 10^{-15}$ of the cluster. As this is a very small velocity, producing a displacement of only $400$ km in the age of the Universe, we conclude that our \frost{} code is essentially momentum-conserving for any plausible stellar-dynamical applications.}
\label{fig: energy-million}
\end{figure*}

We generate realistic million-body ($N=10^6$) star cluster initial conditions for our \frost{} simulations using the McLuster code \citep{Kupper2011}. We use the common density profile of \cite{Plummer1911}, and the mass distribution of the stellar population corresponds to an initial mass function of \cite{Kroupa2001} evolved to an age of $1$ Gyr after which the masses of the stars and compact remnants range from $0.08\,M_\mathrm{\odot}$ to $\sim11\,M_\mathrm{\odot}$. The half-mass radius of the cluster is $r_\mathrm{1/2} = 3.5\,$pc and its total stellar mass is $M=3.91\times10^5\,M_\odot$, i.e. the cluster model is somewhat more massive than an average Milky Way globular cluster \citep{Heggie2003}. For additional details about the star cluster initial conditions see Appendix \ref{appendix: plummer}.

We run the million-body initial conditions using \frost{} for $50$ N-body time units ($t=50\,t_\mathrm{nb}$) of the star cluster corresponding to approximately $12$ Myr of simulation time \citep{Heggie1986}. The integration accuracy parameter is set to $\eta = 0.2$. We test two different values of gravitational softening in two separate simulation runs. In the first run we use a gravitational softening of $\epsilon_\mathrm{P} = 10^{-3}$ pc while in another simulation the softening parameter is set to an extremely small value of $\epsilon_\mathrm{P} = 1\,R_\mathrm{\odot} \sim 2.3\times10^{-8}$ pc. The values of total energy $E$, momentum $P=\norm{\vect{P}}$ and angular momentum $L=\norm{\vect{L}}$ of the cluster are measured every $0.01$ Myr during the simulation.

The time evolution of the relative energy error $|\Delta E/E|$, the relative angular momentum error $|\Delta L/L|$ and the linear momentum error $|\Delta P|$ is displayed in Fig. \ref{fig: energy-million}. The momentum error is presented as $|\Delta P|/M$ i.e. the (initially zero) center-of-mass velocity of the cluster in the units of km/s. The chosen gravitational softening parameter has no apparent effect on the conservation of the three studied quantities. Beginning from the left panel Fig. \ref{fig: energy-million} we see that the relative energy error is initially $|\Delta E/E|\sim 10^{-11}$ and $|\Delta E/E|\sim 2\times10^{-10}$ at the end of the simulations. The energy error does not increase at a constant rate but in brief intervals among longer periods without considerable error growth. This energy error behaviour is a manifestation of the fact that no integration method with discretised time-steps can be made perfectly time-symmetric \citep{Dehnen2017b}. As the symmetrised time-steps of Eq. \eqref{eq: time-step-factor} restore the time-reversibility only approximately \citep{Pelupessy2012} some error growth is inevitable.

The middle and the right panels of Fig. \ref{fig: energy-million} show the evolution of the relative angular momentum error $|\Delta L/L|$ and the absolute linear momentum $|\Delta P|/M$ in the units of center-of-mass velocity. Both of the quantities remain very close to a constant value during the entire simulation time. The relative angular momentum error is approximately $|\Delta L/L| \sim 10^{-13}$. The center-of-mass velocity is of the order of $|\Delta P|/M \sim 10^{-15}$ i.e. $|\Delta P|\sim 10^{-10}$. We emphasise that a center-of-mass velocity of the order of $|\Delta P|/M \sim 10^{-15}$ km/s corresponds to a center-of-mass displacement of only $\sim 400$ km over the age of the Universe. 

We conclude that our \frost{} code is essentially momentum-conserving and conserves energy well in all stellar-dynamical applications examined in this study. However, we note that reaching similar accuracy in more extreme simulation setups such as gigayear-long integrations in which star clusters evolve beyond the core collapse \citep{Konstantinidis2010,Pelupessy2012,Wang2016} requires a special treatment of binaries and close particle encounters which our code does not yet include. We briefly discuss the implementation options for these algorithms in Section \ref{section: 7}.

\subsection{Scaling experiments}

\begin{table}
	\centering
	\caption{The properties of the five star cluster models used in this study. Each cluster has $r_\mathrm{1/2} = 3.5$ pc and $t_\mathrm{age} = 1$ Gyr.}
	\label{tab: clusters}
	\begin{tabular}{cccc}
		\hline
		Cluster & $N$ & $M$ $[M_\odot]$ & $t_\mathrm{nb}$ $[$Myr$]$\\
		\hline
		A & $1.00\times10^5$ & $3.79\times10^4$ & $0.74$\\
		B & $3.16\times10^5$ & $1.22\times10^5$ & $0.41$\\
		C & $1.00\times10^6$ & $3.91\times10^5$ & $0.23$\\
		D & $3.16\times10^6$ & $1.25\times10^6$ & $0.13$\\
		E & $1.00\times10^7$ & $3.93\times10^6$ & $0.07$\\
		\hline
	\end{tabular}
\end{table}

\begin{figure*}
\includegraphics[width=\textwidth]{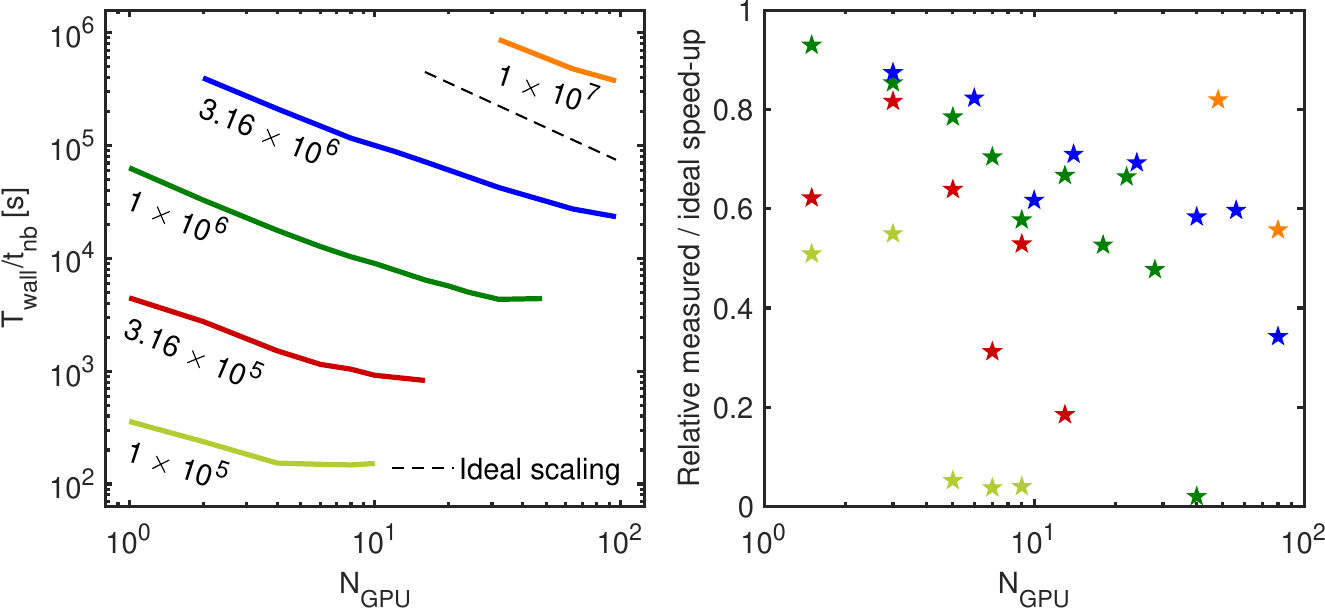}
\caption{Left panel: The measured (solid lines) strong scaling behaviour of the \frost{} code in a series of scaling experiments with star cluster models A-E of Table \ref{tab: clusters}. The code scales linearly (the dashed line) until the scaling stalls around $N_\mathrm{GPU}^\mathrm{max} \approx 4\times N/10^5$, where $N$ is the number of particles in the cluster. This implies that \frost{} scales well up to $N_\mathrm{GPU} \sim 400$ in simulations with $N=10^7$ particles. A million-body run requires approximately one hour of wall-clock time per N-body timescale using $N_\mathrm{GPU} \sim 30$--$40$. The numerical accuracy of the runs was very high, $|\Delta E/E| \sim 10^{-10}$. Reducing the numerical accuracy by increasing the $\eta$ parameter would speed up the million-body run. Right panel: the relative speed-up of the code when increasing $N_\mathrm{GPU}$ compared to the ideal linear scaling. The code scaling is closer to ideal with high particle numbers $N\gtrsim10^6$ and with lower number of GPUs, as expected.}
\label{fig: scaling}
\end{figure*}
Finally we run a set of timing tests in order to study the scaling of the \frost{} code. We generate four additional stellar cluster models with the recipe presented in Section \ref{section: million-conservation} and Appendix \ref{appendix: plummer}. The smallest cluster consists of $N=1.00\times10^5$ particles while the most massive cluster model has $N=1.00\times10^7$ particles. The logarithms of the particle numbers of the five star clusters are linearly spaced yielding an expected tenfold increase in the simulation wall-clock time when comparing a cluster to the next largest one. The relevant physical properties of the cluster models are listed in Table \ref{tab: clusters}. The integrator accuracy parameter was set to $\eta = 0.2$ and the gravitational softening to $\epsilon_\mathrm{P} = 10^{-3}$ pc.

The scaling tests in this study measure the strong scaling of the \frost{} code as we keep the problem size fixed while increasing the amount of computational resources. We always use $N_\mathrm{GPU} = N_\mathrm{CPU}$. The maximum number of GPUs employed was $N_\mathrm{GPU} = 96$. The scaling experiments were performed using the MPG supercomputer Cobra of the Max Planck Computing and Data facility (MPCDF). At the time when the \frost{} scaling experiments were performed each hardware-accelerated Cobra node hosted two Nvidia Tesla V100-PCIE-32GB GPUs.
 
The results of the \frost{} scaling experiments are displayed in Fig. \ref{fig: scaling}. The figure shows the elapsed wall-clock time per N-body timescale $T_\mathrm{wall}/t_\mathrm{nb}$ as a function of the number of GPUs ($N_\mathrm{GPU}$). The numerical accuracy of each simulation was comparable to the run presented in Fig. \ref{fig: energy-million}. Starting from the results of the smallest cluster model A with $N=10^5$ particles we find that the code scales linearly until $N_\mathrm{GPU} \sim 4$ after which the scaling stalls. This happens as the particle number per GPU decreases and becomes smaller than the number of concurrent threads on the GPUs. With 4 GPUs running the cluster model A with \frost{} for a single N-body time takes approximately a few minutes.

The run with the cluster model C with a million stars is approximately the modern state-of-the-art size of a direct-summation simulation. With \frost{} the required wall-clock time to run this simulation for one $t_\mathrm{nb}$ is close to an hour. Brief parameter tests show that reducing the numerical accuracy by increasing the $\eta$ parameter to $\eta = 0.8$ speeds up the million-body run to $\sim20$ minutes per $t_\mathrm{nb}$. In this case the relative energy error is $\sim{10^{-8}}$.

We find that the scaling of the \frost{} code stalls when the number of GPUs reaches approximately $N_\mathrm{GPU}^\mathrm{max}$ defined as
\begin{equation}
    N_\mathrm{GPU}^\mathrm{max} \approx 4\times \frac{N}{10^5}
\end{equation}
in which $N$ is the number of simulation particles. This empirical relation suggests that \frost{} scales until $N_\mathrm{GPU}^\mathrm{max} \sim 400$ GPUs with $N=10^7$ simulation particles. However, we do not perform the scaling tests beyond $N_\mathrm{GPU}=100$ in this study due to the limited number of GPU-accelerated nodes on the Cobra supercomputer and such simulations will be included in future work. 

Finally we estimate the running times for simulations using $N_\mathrm{GPU} = N_\mathrm{GPU}^\mathrm{max}$ GPUs. For $N=10^6$ simulation particles with $\eta = 0.4$ we expect $T_\mathrm{wall} \sim 2$ weeks per Gyr as doubling the accuracy parameter $\eta$ increases the wall-clock time by a factor of two. Going beyond million-particle runs with the same integration accuracy parameter, the $N=5\times10^6$ run yields approximately $T_\mathrm{wall} \sim 4$ weeks per $100$ Myr and $T_\mathrm{wall} \sim 4$ weeks per $10$ Myr for $N=10^7$ particles. With even higher values of $\eta$ would further speed up the code at the cost of decreased numerical accuracy.

The star cluster models in this study did not include primordial stellar binaries as our code does not yet include special integration techniques for binaries and close particle encounters. In general primordial binaries increase the running times of the codes especially when the fraction of binary stars is high. The exact increase of the run time highly depends on the numerical implementation of the simulation code and the initial conditions. For the widely used simulation code \texttt{NBODY6++GPU}, including $5\%$ of primordial binaries in a million-body simulation increases the running time by a factor of $\sim 2$ due to the use of a serial KS regularization method for binaries \citep{Wang2015}. The recent \texttt{PeTar} code \citep{Wang2020b} can treat arbitrary binary fractions with a parallelised regularization method SDAR \citep{Wang2020a}, providing a speed-up of approximately an order of magnitude compared to serial regularisation methods. As a parallel treatment of binaries is the key to simulating large binary fractions, the future regularisation algorithm for binaries in \frost{} will be the modern highly parallelised MSTAR algorithm written by the authors \citep{Rantala2020}. We expect that running simulations with large binary fractions will be up to a factor of a few more expensive than the \frost{} simulations described above.


\section{Summary and Conclusions}\label{section: 7}

In this study we have derived a novel hierarchical generalisation of the fourth-order forward symplectic integrator. The HHS-FSI integrator implemented in the new direct N-body simulation code \frost{} has several desirable properties as described below.

\begin{itemize}
    \item The integrator is very suitable for problems with \emph{an extremely large dynamical range} due to the use of hierarchical Hamiltonian splitting which essentially decouples the evolution of the rapidly evolving parts of the system from the slowly evolving regions.
    \item \emph{The integrator is of the fourth order.} This fact allows for obtaining more accurate simulation results than with a second-order symplectic integrator in similar wall-clock time or equally accurate simulation results faster.
    \item The integrator uses \emph{strictly positive (i.e. forward) time-steps} unlike other high-order symplectic integrators \citep{Yoshida1990}. Forward integrators have been show to be more accurate than their counterparts including negative time-steps, at least for few-body problems \citep{Chin2007}. In addition, negative time-steps may considerably reduce the efficiency of hierarchical integrators \citep{Pelupessy2012} which our integrator completely avoids.
    \item \emph{The integrator is symplectic} i.e. there is no secular energy error growth in long-term simulations unlike many widely-used fourth-order integrators (e.g. \citealt{Aarseth2003,Binney2008}). However, this statement is strictly true only with constant time-steps which can be efficiently used if the particle number is somewhat low. Thus, we use individual adaptive time-steps to reach high $N\gtrsim10^6$ particle numbers at the cost of formal time-reversibility (and thus symplecticity) of our integrator. We approximately restore the lost time-reversibility of our integrator by introducing the so-called time-step symmetrised procedure  \citep{Pelupessy2012,Dehnen2017b}. This procedure limits the secular energy drift in simulations to manageable levels and allows for accurate long-term simulation runs.
\end{itemize}
We have implemented the novel integration method into an integrator code package \frost{}. The code is written in MPI-parallelised CUDA C in order to be able to utilise the hardware-accelerated CPU-GPU nodes of the constantly upgrading modern computing clusters and supercomputers. We have so far tested the \frost{} code up to $96$ GPUs. We provide implementation instructions for most important functions of \frost{} in a pseudocode format to ease the numerical implementation of future hierarchical fourth-order forward integrators by the numerical astrophysics community.

We have verified the numerical accuracy of the \frost{} code in both few-body and million-body regime. The results of the few-body experiments with Keplerian binaries and Solar system analogues confirm that our integrator implementation is indeed of the fourth order. The minimum relative energy error in the simulations is $|\Delta E/E| \sim 10^{-13}$--$10^{-10}$ depending on the eccentricity of the two-body orbital elements of the particles in the initial conditions. Linear and angular momentum are conserved up to the noise floor set by the floating-point round-off error, for linear momentum $|\Delta P/P| \lesssim 10^{-15}$ and $|\Delta L/L| \lesssim 10^{-13}$. The effect of round-off error increases towards smaller integration accuracy parameters $\eta$, as expected. In simulations with stellar cluster models containing $N=10^6$ single stars we find that the code reaches the accuracy of $|\Delta E/E| \sim 10^{-10}$ regardless of the gravitational softening used. In these runs angular momentum error remains constant at $|\Delta L/L| \sim 10^{-13}$ while the linear momentum error corresponds to a center-of-mass displacement of only a few hundred kilometers for the star cluster in the age of the Universe.

We performed a set of simulations with particle numbers $10^5 \leq N \leq 10^7$ and up to approximately a hundred GPUs in order to measure the strong scaling of the \frost{} code. The code scales with small number of GPUs almost ideally after which the scaling is still linear, though deviates from the ideal scaling law. The scaling tests performed up to $N_\mathrm{GPU}$ indicate that the scaling of \frost{} stalls approximately at $N_\mathrm{GPU}^\mathrm{max} \approx 4\times N/10^5$ GPUs. The observed scaling behaviour of the code indicates that simulations with $N=5\times10^6$ to $N=10^7$ could be run using $N_\mathrm{GPU} \sim 200$--$400$. Due to its good scaling behaviour \frost{} also paves the way towards extended million-body studies of globular clusters and low mass nuclear star clusters with their intermediate-mass black holes on the upcoming next-generation Tier-0 GPU systems like JEWELS booster with several thousand GPUs.

The current code version of \frost{} treats particles as point masses and does not yet include stellar evolution \citep{Hurley2000,Aarseth2003,Wang2015}, collisions and mergers or additional specialised integration recipes for close binary systems. In close binaries (possibly dissipative) forces beyond Newtonian gravity may become important. Important examples of such cases are relativistic post-Newtonian corrections (e.g. \citealt{Poisson2014} and references therein),
or binary stellar evolution phenomena such as mass transfer (e.g. \citealt{Hurley2002}) and tides (e.g. \citealt{Mardling2001,Samsing2018}). The further spatial Hamiltonian splitting of the individual hierarchy levels into field stars, binaries and multiple star systems allows for straightforward inclusion of specialised external integration modules into \frost{} in future work. These modules, such as regularised integrators (e.g. \citealt{Mikkola2006,Mikkola2008,Rantala2017,Rantala2020,Wang2020a}), Wisdom-Holman integrators and Kepler solvers (e.g. \citealt{Wisdom1991,Wisdom2015,Rein2015,Dehnen2017a}) or secular multiple star evolution codes (e.g. \citealt{Hamers2016, Hamers2020}) can be used when extreme numerical precision or computational speed (or both) are required for few-body systems in the fastest levels of the time-step hierarchy.

Finally, one may wonder whether even higher-order generalisations of the presented hierarchical fourth-order forward integrator exist. Unfortunately, forward symplectic integrators of the order six have not been discovered while the proof of their possible non-existence also remains elusive \citep{Chin2005}. Another complication in possible future higher-order forward symplectic integrators is the increasing complexity of the nested commutator terms required for the algorithm (e.g. \citealt{Dehnen2017a}). It is unlikely that such terms can be evaluated in a straightforward manner, most probably preventing the construction of a practical forward integrator (hierarchical or not) beyond the fourth order.

\section*{Data availability statement}
The relevant initial conditions and the data presented in \Cref{fig: binary,fig: solarsystem,fig: energy-million,fig: scaling} of this article will be shared on reasonable request to the corresponding author.

\section*{Acknowledgements}
The authors thank the anonymous referee for a constructive review process. We also thank Walter Dehnen and Long Wang for valuable comments on the manuscript. The numerical simulations were performed using facilities hosted by the Max Planck Computing and Data Facility (MPCDF) and the Leibniz Supercomputing Centre (LRZ), Germany. TN acknowledges support from the Deutsche Forschungsgemeinschaft (DFG, German Research Foundation) under Germany's Excellence Strategy - EXC-2094 - 390783311 from the DFG Cluster of Excellence "ORIGINS".


\bibliographystyle{mnras}
\interlinepenalty=10000
\bibliography{references}

\begin{thebibliography}{}
\makeatletter
\relax
\def\mn@urlcharsother{\let\do\@makeother \do\$\do\&\do\#\do\^\do\_\do\%\do\~}
\def\mn@doi{\begingroup\mn@urlcharsother \@ifnextchar [ {\mn@doi@}
  {\mn@doi@[]}}
\def\mn@doi@[#1]#2{\def\@tempa{#1}\ifx\@tempa\@empty \href
  {http://dx.doi.org/#2} {doi:#2}\else \href {http://dx.doi.org/#2} {#1}\fi
  \endgroup}
\def\mn@eprint#1#2{\mn@eprint@#1:#2::\@nil}
\def\mn@eprint@arXiv#1{\href {http://arxiv.org/abs/#1} {{\tt arXiv:#1}}}
\def\mn@eprint@dblp#1{\href {http://dblp.uni-trier.de/rec/bibtex/#1.xml}
  {dblp:#1}}
\def\mn@eprint@#1:#2:#3:#4\@nil{\def\@tempa {#1}\def\@tempb {#2}\def\@tempc
  {#3}\ifx \@tempc \@empty \let \@tempc \@tempb \let \@tempb \@tempa \fi \ifx
  \@tempb \@empty \def\@tempb {arXiv}\fi \@ifundefined
  {mn@eprint@\@tempb}{\@tempb:\@tempc}{\expandafter \expandafter \csname
  mn@eprint@\@tempb\endcsname \expandafter{\@tempc}}}

\bibitem[\protect\citeauthoryear{{Aarseth}}{{Aarseth}}{1999}]{Aarseth1999}
{Aarseth} S.~J.,  1999, \mn@doi [\pasp] {10.1086/316455}, \href
  {https://ui.adsabs.harvard.edu/abs/1999PASP..111.1333A} {111, 1333}

\bibitem[\protect\citeauthoryear{{Aarseth}}{{Aarseth}}{2003}]{Aarseth2003}
{Aarseth} S.~J.,  2003, {Gravitational N-Body Simulations}.
Cambridge University Press

\bibitem[\protect\citeauthoryear{{Aguilar-Arg{\"u}ello}, {Valenzuela},
  {Clemente}, {Vel{\'a}zquez}  \& {Trelles}}{{Aguilar-Arg{\"u}ello}
  et~al.}{2020}]{AguilarArguello2020}
{Aguilar-Arg{\"u}ello} G.,  {Valenzuela} O.,  {Clemente} J.~C.,
  {Vel{\'a}zquez} H.,   {Trelles} J.~A.,  2020, arXiv e-prints, \href
  {https://ui.adsabs.harvard.edu/abs/2020arXiv200906133A} {p. arXiv:2009.06133}

\bibitem[\protect\citeauthoryear{{Ahmad} \& {Cohen}}{{Ahmad} \&
  {Cohen}}{1973}]{Ahmad1973}
{Ahmad} A.,  {Cohen} L.,  1973, \mn@doi [Journal of Computational Physics]
  {10.1016/0021-9991(73)90160-5}, \href
  {https://ui.adsabs.harvard.edu/abs/1973JCoPh..12..389A} {12, 389}

\bibitem[\protect\citeauthoryear{Baker}{Baker}{1902}]{Baker1902}
Baker H.~F.,  1902, \mn@doi [Proceedings of the London Mathematical Society]
  {https://doi.org/10.1112/plms/s1-35.1.333}, s1-35, 333

\bibitem[\protect\citeauthoryear{Baker}{Baker}{1905}]{Baker1905}
Baker H.~F.,  1905, \mn@doi [Proceedings of the London Mathematical Society]
  {https://doi.org/10.1112/plms/s2-3.1.24}, s2-3, 24

\bibitem[\protect\citeauthoryear{{Barnes}}{{Barnes}}{2012}]{Barnes2012}
{Barnes} J.~E.,  2012, \mn@doi [\mnras] {10.1111/j.1365-2966.2012.21462.x},
  \href {https://ui.adsabs.harvard.edu/abs/2012MNRAS.425.1104B} {425, 1104}

\bibitem[\protect\citeauthoryear{{Binney} \& {Tremaine}}{{Binney} \&
  {Tremaine}}{2008}]{Binney2008}
{Binney} J.,  {Tremaine} S.,  2008, {Galactic Dynamics: Second Edition}.
Princeton University Press, Princeton, NJ USA

\bibitem[\protect\citeauthoryear{Campbell}{Campbell}{1896}]{Campbell1896}
Campbell J.~E.,  1896, \mn@doi [Proceedings of the London Mathematical Society]
  {https://doi.org/10.1112/plms/s1-28.1.381}, s1-28, 381

\bibitem[\protect\citeauthoryear{Campbell}{Campbell}{1897}]{Campbell1897}
Campbell J.~E.,  1897, \mn@doi [Proceedings of the London Mathematical Society]
  {https://doi.org/10.1112/plms/s1-29.1.14}, s1-29, 14

\bibitem[\protect\citeauthoryear{{Chin}}{{Chin}}{1997}]{Chin1997}
{Chin} S.~A.,  1997, \mn@doi [Physics Letters A]
  {https://doi.org/10.1016/S0375-9601(97)00003-0}, 226, 344

\bibitem[\protect\citeauthoryear{{Chin}}{{Chin}}{2007a}]{Chin2007}
{Chin} S.~A.,  2007a, arXiv e-prints, p. arXiv:0704.3273

\bibitem[\protect\citeauthoryear{{Chin}}{{Chin}}{2007b}]{Chin2007a}
{Chin} S.~A.,  2007b, \mn@doi [\pre] {10.1103/PhysRevE.75.036701}, 75, 036701

\bibitem[\protect\citeauthoryear{{Chin} \& {Chen}}{{Chin} \&
  {Chen}}{2005}]{Chin2005}
{Chin} S.~A.,  {Chen} C.~R.,  2005, \mn@doi [Celestial Mechanics and Dynamical
  Astronomy] {10.1007/s10569-004-4622-z}, 91, 301

\bibitem[\protect\citeauthoryear{{Danby}}{{Danby}}{1992}]{Danby1992}
{Danby} J. M.~A.,  1992, {Fundamentals of celestial mechanics}.
Willmann-Bell, Richmond, Va., U.S.

\bibitem[\protect\citeauthoryear{{Dehnen}}{{Dehnen}}{2017}]{Dehnen2017b}
{Dehnen} W.,  2017, \mn@doi [\mnras] {10.1093/mnras/stx1944}, \href
  {https://ui.adsabs.harvard.edu/abs/2017MNRAS.472.1226D} {472, 1226}

\bibitem[\protect\citeauthoryear{{Dehnen} \& {Hernandez}}{{Dehnen} \&
  {Hernandez}}{2017}]{Dehnen2017a}
{Dehnen} W.,  {Hernandez} D.~M.,  2017, \mn@doi [\mnras]
  {10.1093/mnras/stw2758}, \href
  {https://ui.adsabs.harvard.edu/abs/2017MNRAS.465.1201D} {465, 1201}

\bibitem[\protect\citeauthoryear{{Dehnen} \& {Read}}{{Dehnen} \&
  {Read}}{2011}]{Dehnen2011}
{Dehnen} W.,  {Read} J.~I.,  2011, \mn@doi [European Physical Journal Plus]
  {10.1140/epjp/i2011-11055-3}, \href
  {https://ui.adsabs.harvard.edu/abs/2011EPJP..126...55D} {126, 55}

\bibitem[\protect\citeauthoryear{{Dragt} \& {Finn}}{{Dragt} \&
  {Finn}}{1976}]{Dragt1976}
{Dragt} A.~J.,  {Finn} J.~M.,  1976, \mn@doi [Journal of Mathematical Physics]
  {10.1063/1.522868}, \href
  {https://ui.adsabs.harvard.edu/abs/1976JMP....17.2215D} {17, 2215}

\bibitem[\protect\citeauthoryear{{Farr} \& {Bertschinger}}{{Farr} \&
  {Bertschinger}}{2007}]{Farr2007}
{Farr} W.~M.,  {Bertschinger} E.,  2007, \mn@doi [\apj] {10.1086/518641}, \href
  {https://ui.adsabs.harvard.edu/abs/2007ApJ...663.1420F} {663, 1420}

\bibitem[\protect\citeauthoryear{{Gaburov}, {Harfst}  \& {Portegies
  Zwart}}{{Gaburov} et~al.}{2009}]{Gaburov2009}
{Gaburov} E.,  {Harfst} S.,   {Portegies Zwart} S.,  2009, \na, 14, 630

\bibitem[\protect\citeauthoryear{Goldman \& Kaper}{Goldman \&
  Kaper}{1996}]{Goldman1996}
Goldman D.,  Kaper T.~J.,  1996, SIAM Journal on Numerical Analysis, 33, 349

\bibitem[\protect\citeauthoryear{Goldstein}{Goldstein}{1980}]{Goldstein1980}
Goldstein H.,  1980, Classical Mechanics.
Addison-Wesley

\bibitem[\protect\citeauthoryear{Hairer, Lubich  \& Wanner}{Hairer
  et~al.}{2006}]{Hairer2006}
Hairer E.,  Lubich C.,   Wanner G.,  2006, Geometric numerical integration:
  structure-preserving algorithms for ordinary differential equations.
~ Vol. 31, Springer-Verlag, Berlin Heidelberg

\bibitem[\protect\citeauthoryear{{Hamers} \& {Portegies Zwart}}{{Hamers} \&
  {Portegies Zwart}}{2016}]{Hamers2016}
{Hamers} A.~S.,  {Portegies Zwart} S.~F.,  2016, \mn@doi [\mnras]
  {10.1093/mnras/stw784}, \href
  {https://ui.adsabs.harvard.edu/abs/2016MNRAS.459.2827H} {459, 2827}

\bibitem[\protect\citeauthoryear{{Hamers}, {Rantala}, {Neunteufel}, {Preece}
  \& {Vynatheya}}{{Hamers} et~al.}{2020}]{Hamers2020}
{Hamers} A.~S.,  {Rantala} A.,  {Neunteufel} P.,  {Preece} H.,   {Vynatheya}
  P.,  2020, arXiv e-prints, \href
  {https://ui.adsabs.harvard.edu/abs/2020arXiv201104513H} {p. arXiv:2011.04513}

\bibitem[\protect\citeauthoryear{{Hands}, {Dehnen}, {Gration}, {Stadel}  \&
  {Moore}}{{Hands} et~al.}{2019}]{Hands2019}
{Hands} T.~O.,  {Dehnen} W.,  {Gration} A.,  {Stadel} J.,   {Moore} B.,  2019,
  \mn@doi [\mnras] {10.1093/mnras/stz1069}, \href
  {https://ui.adsabs.harvard.edu/abs/2019MNRAS.490...21H} {490, 21}

\bibitem[\protect\citeauthoryear{{Hausdorff}}{{Hausdorff}}{1906}]{Hausdorff1906}
{Hausdorff} F.,  1906, {Ber. über die Verhandlungen der Königl. Sächs. Ges.
  der Wiss. zu Leipzig. Math.-phys.}, 58, 19

\bibitem[\protect\citeauthoryear{{Heggie} \& {Hut}}{{Heggie} \&
  {Hut}}{2003}]{Heggie2003}
{Heggie} D.,  {Hut} P.,  2003, {The Gravitational Million-Body Problem: A
  Multidisciplinary Approach to Star Cluster Dynamics}.
Cambridge University Press

\bibitem[\protect\citeauthoryear{{Heggie} \& {Mathieu}}{{Heggie} \&
  {Mathieu}}{1986}]{Heggie1986}
{Heggie} D.~C.,  {Mathieu} R.~D.,  1986, in {Hut} P.,  {McMillan} S. L.~W.,
  eds, , Vol.~267, The Use of Supercomputers in Stellar Dynamics.
Springer-Verlag, Berlin Heidelberg New York, p.~233,
  \mn@doi{10.1007/BFb0116419}

\bibitem[\protect\citeauthoryear{{Hernandez} \& {Bertschinger}}{{Hernandez} \&
  {Bertschinger}}{2015}]{Hernandez2015}
{Hernandez} D.~M.,  {Bertschinger} E.,  2015, \mn@doi [\mnras]
  {10.1093/mnras/stv1439}, \href
  {https://ui.adsabs.harvard.edu/abs/2015MNRAS.452.1934H} {452, 1934}

\bibitem[\protect\citeauthoryear{{Hernandez} \& {Bertschinger}}{{Hernandez} \&
  {Bertschinger}}{2018}]{Hernandez2018}
{Hernandez} D.~M.,  {Bertschinger} E.,  2018, \mn@doi [\mnras]
  {10.1093/mnras/sty184}, \href
  {https://ui.adsabs.harvard.edu/abs/2018MNRAS.475.5570H} {475, 5570}

\bibitem[\protect\citeauthoryear{{Hernandez} \& {Holman}}{{Hernandez} \&
  {Holman}}{2020}]{Hernandez2020}
{Hernandez} D.~M.,  {Holman} M.~J.,  2020, arXiv e-prints, \href
  {https://ui.adsabs.harvard.edu/abs/2020arXiv201013907H} {p. arXiv:2010.13907}

\bibitem[\protect\citeauthoryear{{Hernquist} \& {Katz}}{{Hernquist} \&
  {Katz}}{1989}]{Hernquist1989}
{Hernquist} L.,  {Katz} N.,  1989, \mn@doi [\apjs] {10.1086/191344}, \href
  {https://ui.adsabs.harvard.edu/abs/1989ApJS...70..419H} {70, 419}

\bibitem[\protect\citeauthoryear{Holder, Leimkuhler  \& Reich}{Holder
  et~al.}{1999}]{Holder1999}
Holder T.,  Leimkuhler B.,   Reich S.,  1999, Appl. Numer. Math, 39, 367

\bibitem[\protect\citeauthoryear{{Hubber}, {Rosotti}  \& {Booth}}{{Hubber}
  et~al.}{2018}]{Hubber2018}
{Hubber} D.~A.,  {Rosotti} G.~P.,   {Booth} R.~A.,  2018, \mn@doi [\mnras]
  {10.1093/mnras/stx2405}, \href
  {https://ui.adsabs.harvard.edu/abs/2018MNRAS.473.1603H} {473, 1603}

\bibitem[\protect\citeauthoryear{{Hurley}, {Pols}  \& {Tout}}{{Hurley}
  et~al.}{2000}]{Hurley2000}
{Hurley} J.~R.,  {Pols} O.~R.,   {Tout} C.~A.,  2000, \mn@doi [\mnras]
  {10.1046/j.1365-8711.2000.03426.x}, \href
  {https://ui.adsabs.harvard.edu/abs/2000MNRAS.315..543H} {315, 543}

\bibitem[\protect\citeauthoryear{{Hurley}, {Tout}  \& {Pols}}{{Hurley}
  et~al.}{2002}]{Hurley2002}
{Hurley} J.~R.,  {Tout} C.~A.,   {Pols} O.~R.,  2002, \mn@doi [\mnras]
  {10.1046/j.1365-8711.2002.05038.x}, \href
  {https://ui.adsabs.harvard.edu/abs/2002MNRAS.329..897H} {329, 897}

\bibitem[\protect\citeauthoryear{{Hut}, {Makino}  \& {McMillan}}{{Hut}
  et~al.}{1995}]{Hut1995}
{Hut} P.,  {Makino} J.,   {McMillan} S.,  1995, \mn@doi [\apjl]
  {10.1086/187844}, \href
  {https://ui.adsabs.harvard.edu/abs/1995ApJ...443L..93H} {443, L93}

\bibitem[\protect\citeauthoryear{{Ito}, {Makino}, {Ebisuzaki}  \&
  {Sugimoto}}{{Ito} et~al.}{1990}]{Ito1990}
{Ito} T.,  {Makino} J.,  {Ebisuzaki} T.,   {Sugimoto} D.,  1990, \mn@doi
  [Computer Physics Communications] {10.1016/0010-4655(90)90003-J}, \href
  {https://ui.adsabs.harvard.edu/abs/1990CoPhC..60..187I} {60, 187}

\bibitem[\protect\citeauthoryear{{J{\"a}nes}, {Pelupessy}  \& {Portegies
  Zwart}}{{J{\"a}nes} et~al.}{2014}]{Janes2014}
{J{\"a}nes} J.,  {Pelupessy} I.,   {Portegies Zwart} S.,  2014, \mn@doi [\aap]
  {10.1051/0004-6361/201423831}, \href
  {https://ui.adsabs.harvard.edu/abs/2014A&A...570A..20J} {570, A20}

\bibitem[\protect\citeauthoryear{{Konstantinidis} \&
  {Kokkotas}}{{Konstantinidis} \& {Kokkotas}}{2010}]{Konstantinidis2010}
{Konstantinidis} S.,  {Kokkotas} K.~D.,  2010, \mn@doi [\aap]
  {10.1051/0004-6361/200913890}, \href
  {https://ui.adsabs.harvard.edu/abs/2010A&A...522A..70K} {522, A70}

\bibitem[\protect\citeauthoryear{{Kroupa}}{{Kroupa}}{2001}]{Kroupa2001}
{Kroupa} P.,  2001, \mn@doi [\mnras] {10.1046/j.1365-8711.2001.04022.x}, \href
  {https://ui.adsabs.harvard.edu/abs/2001MNRAS.322..231K} {322, 231}

\bibitem[\protect\citeauthoryear{{K{\"u}pper}, {Maschberger}, {Kroupa}  \&
  {Baumgardt}}{{K{\"u}pper} et~al.}{2011}]{Kupper2011}
{K{\"u}pper} A. H.~W.,  {Maschberger} T.,  {Kroupa} P.,   {Baumgardt} H.,
  2011, \mn@doi [\mnras] {10.1111/j.1365-2966.2011.19412.x}, \href
  {https://ui.adsabs.harvard.edu/abs/2011MNRAS.417.2300K} {417, 2300}

\bibitem[\protect\citeauthoryear{{Kustaanheimo} \& {Stiefel}}{{Kustaanheimo} \&
  {Stiefel}}{1965}]{Kustaanheimo1965}
{Kustaanheimo} P.,  {Stiefel} E.,  1965, J. Reine Angew. Math, 218, 204

\bibitem[\protect\citeauthoryear{{Makino}}{{Makino}}{1991}]{Makino1991}
{Makino} J.,  1991, \pasj, \href
  {https://ui.adsabs.harvard.edu/abs/1991PASJ...43..859M} {43, 859}

\bibitem[\protect\citeauthoryear{{Makino}}{{Makino}}{2008}]{Makino2008}
{Makino} J.,  2008, in {Vesperini} E.,  {Giersz} M.,   {Sills} A.,  eds,  IAU
  Symposium Vol. 246, Dynamical Evolution of Dense Stellar Systems. pp
  457--466, \mn@doi{10.1017/S1743921308016165}

\bibitem[\protect\citeauthoryear{{Makino} \& {Aarseth}}{{Makino} \&
  {Aarseth}}{1992}]{Makino1992}
{Makino} J.,  {Aarseth} S.~J.,  1992, \pasj, \href
  {https://ui.adsabs.harvard.edu/abs/1992PASJ...44..141M} {44, 141}

\bibitem[\protect\citeauthoryear{{Makino}, {Hut}, {Kaplan}  \&
  {Sayg{\i}n}}{{Makino} et~al.}{2006}]{Makino2006}
{Makino} J.,  {Hut} P.,  {Kaplan} M.,   {Sayg{\i}n} H.,  2006, \mn@doi [\na]
  {10.1016/j.newast.2006.06.003}, \href
  {https://ui.adsabs.harvard.edu/abs/2006NewA...12..124M} {12, 124}

\bibitem[\protect\citeauthoryear{{Mardling} \& {Aarseth}}{{Mardling} \&
  {Aarseth}}{2001}]{Mardling2001}
{Mardling} R.~A.,  {Aarseth} S.~J.,  2001, \mn@doi [\mnras]
  {10.1046/j.1365-8711.2001.03974.x}, \href
  {https://ui.adsabs.harvard.edu/abs/2001MNRAS.321..398M} {321, 398}

\bibitem[\protect\citeauthoryear{McLachlan}{McLachlan}{1995}]{Mclachlan1995}
McLachlan R.~I.,  1995, \mn@doi [SIAM J. Sci. Comput.] {10.1137/0916010}, 16,
  151

\bibitem[\protect\citeauthoryear{{Mikkola}}{{Mikkola}}{2008}]{Mikkola2008a}
{Mikkola} S.,  2008, in {Vesperini} E.,  {Giersz} M.,   {Sills} A.,  eds,  IAU
  Symposium Vol. 246, Dynamical Evolution of Dense Stellar Systems. pp
  218--227, \mn@doi{10.1017/S1743921308015639}

\bibitem[\protect\citeauthoryear{{Mikkola} \& {Aarseth}}{{Mikkola} \&
  {Aarseth}}{1993}]{Mikkola1993}
{Mikkola} S.,  {Aarseth} S.~J.,  1993, \mn@doi [Celestial Mechanics and
  Dynamical Astronomy] {10.1007/BF00695714}, \href
  {https://ui.adsabs.harvard.edu/abs/1993CeMDA..57..439M} {57, 439}

\bibitem[\protect\citeauthoryear{{Mikkola} \& {Merritt}}{{Mikkola} \&
  {Merritt}}{2006}]{Mikkola2006}
{Mikkola} S.,  {Merritt} D.,  2006, \mn@doi [\mnras]
  {10.1111/j.1365-2966.2006.10854.x}, \href
  {https://ui.adsabs.harvard.edu/abs/2006MNRAS.372..219M} {372, 219}

\bibitem[\protect\citeauthoryear{{Mikkola} \& {Merritt}}{{Mikkola} \&
  {Merritt}}{2008}]{Mikkola2008}
{Mikkola} S.,  {Merritt} D.,  2008, \mn@doi [\aj]
  {10.1088/0004-6256/135/6/2398}, \href
  {https://ui.adsabs.harvard.edu/abs/2008AJ....135.2398M} {135, 2398}

\bibitem[\protect\citeauthoryear{{Mikkola} \& {Tanikawa}}{{Mikkola} \&
  {Tanikawa}}{1999}]{Mikkola1999}
{Mikkola} S.,  {Tanikawa} K.,  1999, \mn@doi [\mnras]
  {10.1046/j.1365-8711.1999.02982.x}, \href
  {https://ui.adsabs.harvard.edu/abs/1999MNRAS.310..745M} {310, 745}

\bibitem[\protect\citeauthoryear{{Mukherjee}, {Zhu}, {Trac}  \&
  {Rodriguez}}{{Mukherjee} et~al.}{2020}]{Mukherjee2020}
{Mukherjee} D.,  {Zhu} Q.,  {Trac} H.,   {Rodriguez} C.~L.,  2020, arXiv
  e-prints, \href {https://ui.adsabs.harvard.edu/abs/2020arXiv201202207M} {p.
  arXiv:2012.02207}

\bibitem[\protect\citeauthoryear{{Murray} \& {Dermott}}{{Murray} \&
  {Dermott}}{2000}]{Murray2000}
{Murray} C.~D.,  {Dermott} S.~F.,  2000, {Solar System Dynamics}.
Cambridge University Press

\bibitem[\protect\citeauthoryear{Nguyen}{Nguyen}{2007}]{Nguyen2007}
Nguyen H.,  2007, GPU Gems 3.
Addison-Wesley Professional

\bibitem[\protect\citeauthoryear{{Nitadori} \& {Aarseth}}{{Nitadori} \&
  {Aarseth}}{2012}]{Nitadori2012}
{Nitadori} K.,  {Aarseth} S.~J.,  2012, \mn@doi [\mnras]
  {10.1111/j.1365-2966.2012.21227.x}, \href
  {https://ui.adsabs.harvard.edu/abs/2012MNRAS.424..545N} {424, 545}

\bibitem[\protect\citeauthoryear{{Nitadori} \& {Makino}}{{Nitadori} \&
  {Makino}}{2008}]{Nitadori2008}
{Nitadori} K.,  {Makino} J.,  2008, \mn@doi [\na]
  {10.1016/j.newast.2008.01.010}, \href
  {https://ui.adsabs.harvard.edu/abs/2008NewA...13..498N} {13, 498}

\bibitem[\protect\citeauthoryear{Omelyan}{Omelyan}{2006}]{Omelyan2006}
Omelyan I.,  2006, Physical review. E, Statistical, nonlinear, and soft matter
  physics, 74 3 Pt 2, 036703

\bibitem[\protect\citeauthoryear{{Pelupessy}, {J{\"a}nes}  \& {Portegies
  Zwart}}{{Pelupessy} et~al.}{2012}]{Pelupessy2012}
{Pelupessy} F.~I.,  {J{\"a}nes} J.,   {Portegies Zwart} S.,  2012, \mn@doi
  [\na] {10.1016/j.newast.2012.05.009}, \href
  {https://ui.adsabs.harvard.edu/abs/2012NewA...17..711P} {17, 711}

\bibitem[\protect\citeauthoryear{{Plummer}}{{Plummer}}{1911}]{Plummer1911}
{Plummer} H.~C.,  1911, \mn@doi [\mnras] {10.1093/mnras/71.5.460}, \href
  {https://ui.adsabs.harvard.edu/abs/1911MNRAS..71..460P} {71, 460}

\bibitem[\protect\citeauthoryear{{Poisson} \& {Will}}{{Poisson} \&
  {Will}}{2014}]{Poisson2014}
{Poisson} E.,  {Will} C.~M.,  2014, {Gravity}.
Cambridge University Press

\bibitem[\protect\citeauthoryear{{Preto} \& {Tremaine}}{{Preto} \&
  {Tremaine}}{1999}]{Preto1999}
{Preto} M.,  {Tremaine} S.,  1999, \mn@doi [\aj] {10.1086/301102}, \href
  {https://ui.adsabs.harvard.edu/abs/1999AJ....118.2532P} {118, 2532}

\bibitem[\protect\citeauthoryear{{Rantala}, {Pihajoki}, {Johansson}, {Naab},
  {Lah{\'e}n}  \& {Sawala}}{{Rantala} et~al.}{2017}]{Rantala2017}
{Rantala} A.,  {Pihajoki} P.,  {Johansson} P.~H.,  {Naab} T.,  {Lah{\'e}n} N.,
   {Sawala} T.,  2017, \mn@doi [\apj] {10.3847/1538-4357/aa6d65}, \href
  {https://ui.adsabs.harvard.edu/abs/2017ApJ...840...53R} {840, 53}

\bibitem[\protect\citeauthoryear{{Rantala}, {Pihajoki}, {Mannerkoski},
  {Johansson}  \& {Naab}}{{Rantala} et~al.}{2020}]{Rantala2020}
{Rantala} A.,  {Pihajoki} P.,  {Mannerkoski} M.,  {Johansson} P.~H.,   {Naab}
  T.,  2020, \mn@doi [\mnras] {10.1093/mnras/staa084}, \href
  {https://ui.adsabs.harvard.edu/abs/2020MNRAS.492.4131R} {492, 4131}

\bibitem[\protect\citeauthoryear{{Rein}}{{Rein}}{2020}]{Rein2020}
{Rein} H.,  2020, \mn@doi [\mnras] {10.1093/mnras/staa240}, \href
  {https://ui.adsabs.harvard.edu/abs/2020MNRAS.492.5413R} {492, 5413}

\bibitem[\protect\citeauthoryear{{Rein} \& {Tamayo}}{{Rein} \&
  {Tamayo}}{2015}]{Rein2015}
{Rein} H.,  {Tamayo} D.,  2015, \mn@doi [\mnras] {10.1093/mnras/stv1257}, \href
  {https://ui.adsabs.harvard.edu/abs/2015MNRAS.452..376R} {452, 376}

\bibitem[\protect\citeauthoryear{{Rein}, {Tamayo}  \& {Brown}}{{Rein}
  et~al.}{2019}]{Rein2019}
{Rein} H.,  {Tamayo} D.,   {Brown} G.,  2019, \mn@doi [\mnras]
  {10.1093/mnras/stz2503}, \href
  {https://ui.adsabs.harvard.edu/abs/2019MNRAS.489.4632R} {489, 4632}

\bibitem[\protect\citeauthoryear{{Ruth}}{{Ruth}}{1983}]{Ruth1983}
{Ruth} R.~D.,  1983, \mn@doi [IEEE Transactions on Nuclear Science]
  {10.1109/TNS.1983.4332919}, \href
  {https://ui.adsabs.harvard.edu/abs/1983ITNS...30.2669R} {30, 2669}

\bibitem[\protect\citeauthoryear{{Saha} \& {Tremaine}}{{Saha} \&
  {Tremaine}}{1994}]{Saha1994}
{Saha} P.,  {Tremaine} S.,  1994, \mn@doi [\aj] {10.1086/117210}, \href
  {https://ui.adsabs.harvard.edu/abs/1994AJ....108.1962S} {108, 1962}

\bibitem[\protect\citeauthoryear{{Samsing}, {Leigh}  \& {Trani}}{{Samsing}
  et~al.}{2018}]{Samsing2018}
{Samsing} J.,  {Leigh} N. W.~C.,   {Trani} A.~A.,  2018, \mn@doi [\mnras]
  {10.1093/mnras/sty2247}, \href
  {https://ui.adsabs.harvard.edu/abs/2018MNRAS.481.5436S} {481, 5436}

\bibitem[\protect\citeauthoryear{Sheng}{Sheng}{1989}]{Sheng1989}
Sheng Q.,  1989, \mn@doi [IMA Journal of Numerical Analysis]
  {10.1093/imanum/9.2.199}, 9, 199

\bibitem[\protect\citeauthoryear{{Springel}}{{Springel}}{2005}]{Springel2005}
{Springel} V.,  2005, \mn@doi [\mnras] {10.1111/j.1365-2966.2005.09655.x},
  \href {https://ui.adsabs.harvard.edu/abs/2005MNRAS.364.1105S} {364, 1105}

\bibitem[\protect\citeauthoryear{{Springel}, {Pakmor}, {Zier}  \&
  {Reinecke}}{{Springel} et~al.}{2020}]{Springel2020}
{Springel} V.,  {Pakmor} R.,  {Zier} O.,   {Reinecke} M.,  2020, arXiv
  e-prints, \href {https://ui.adsabs.harvard.edu/abs/2020arXiv201003567S} {p.
  arXiv:2010.03567}

\bibitem[\protect\citeauthoryear{{Suzuki}}{{Suzuki}}{1991}]{Suzuki1991}
{Suzuki} M.,  1991, \mn@doi [Journal of Mathematical Physics]
  {10.1063/1.529425}, \href
  {https://ui.adsabs.harvard.edu/abs/1991JMP....32..400S} {32, 400}

\bibitem[\protect\citeauthoryear{{Suzuki}}{{Suzuki}}{1995}]{Suzuki1995}
{Suzuki} M.,  1995, \mn@doi [Physics Letters A] {10.1016/0375-9601(95)00266-6},
  \href {https://ui.adsabs.harvard.edu/abs/1995PhLA..201..425S} {201, 425}

\bibitem[\protect\citeauthoryear{{Takahashi} \& {Imada}}{{Takahashi} \&
  {Imada}}{1984}]{Takahashi1984}
{Takahashi} M.,  {Imada} M.,  1984, \mn@doi [Journal of the Physical Society of
  Japan] {10.1143/JPSJ.53.3765}, \href
  {https://ui.adsabs.harvard.edu/abs/1984JPSJ...53.3765T} {53, 3765}

\bibitem[\protect\citeauthoryear{{Wang}, {Spurzem}, {Aarseth}, {Nitadori},
  {Berczik}, {Kouwenhoven}  \& {Naab}}{{Wang} et~al.}{2015}]{Wang2015}
{Wang} L.,  {Spurzem} R.,  {Aarseth} S.,  {Nitadori} K.,  {Berczik} P.,
  {Kouwenhoven} M.~B.~N.,   {Naab} T.,  2015, \mn@doi [\mnras]
  {10.1093/mnras/stv817}, \href
  {https://ui.adsabs.harvard.edu/abs/2015MNRAS.450.4070W} {450, 4070}

\bibitem[\protect\citeauthoryear{{Wang} et~al.,}{{Wang}
  et~al.}{2016}]{Wang2016}
{Wang} L.,  et~al., 2016, \mn@doi [\mnras] {10.1093/mnras/stw274}, \href
  {https://ui.adsabs.harvard.edu/abs/2016MNRAS.458.1450W} {458, 1450}

\bibitem[\protect\citeauthoryear{{Wang}, {Nitadori}  \& {Makino}}{{Wang}
  et~al.}{2020a}]{Wang2020a}
{Wang} L.,  {Nitadori} K.,   {Makino} J.,  2020a, \mn@doi [\mnras]
  {10.1093/mnras/staa480}, \href
  {https://ui.adsabs.harvard.edu/abs/2020MNRAS.493.3398W} {493, 3398}

\bibitem[\protect\citeauthoryear{{Wang}, {Iwasawa}, {Nitadori}  \&
  {Makino}}{{Wang} et~al.}{2020b}]{Wang2020b}
{Wang} L.,  {Iwasawa} M.,  {Nitadori} K.,   {Makino} J.,  2020b, \mn@doi
  [\mnras] {10.1093/mnras/staa1915}, \href
  {https://ui.adsabs.harvard.edu/abs/2020MNRAS.497..536W} {497, 536}

\bibitem[\protect\citeauthoryear{{Wisdom} \& {Hernandez}}{{Wisdom} \&
  {Hernandez}}{2015}]{Wisdom2015}
{Wisdom} J.,  {Hernandez} D.~M.,  2015, \mn@doi [\mnras]
  {10.1093/mnras/stv1862}, \href
  {https://ui.adsabs.harvard.edu/abs/2015MNRAS.453.3015W} {453, 3015}

\bibitem[\protect\citeauthoryear{{Wisdom} \& {Holman}}{{Wisdom} \&
  {Holman}}{1991}]{Wisdom1991}
{Wisdom} J.,  {Holman} M.,  1991, \mn@doi [\aj] {10.1086/115978}, \href
  {https://ui.adsabs.harvard.edu/abs/1991AJ....102.1528W} {102, 1528}

\bibitem[\protect\citeauthoryear{{Yoshida}}{{Yoshida}}{1990}]{Yoshida1990}
{Yoshida} H.,  1990, \mn@doi [Physics Letters A]
  {10.1016/0375-9601(90)90092-3}, 150, 262

\bibitem[\protect\citeauthoryear{{Yoshida}}{{Yoshida}}{1993}]{Yoshida1993}
{Yoshida} H.,  1993, \mn@doi [Celestial Mechanics and Dynamical Astronomy]
  {10.1007/BF00699717}, 56, 27

\bibitem[\protect\citeauthoryear{Zhu}{Zhu}{2020}]{Zhu2020}
Zhu Q.,  2020, \mn@doi [New Astronomy]
  {https://doi.org/10.1016/j.newast.2020.101481}, p. 101481

\makeatother
\end{thebibliography}


\appendix

\section{Initial conditions}

\subsection{Solar system with giant planets}\label{appendix: planets}

The numerical integration textbook of \cite{Hairer2006} provides the reference initial conditions for the Sun and the four giant planets in our Solar system. The data originates from "Ahnerts Kalender f\"{u}r Sternfreunde 1994", Johann Ambrosius Barth Verlag 1993, corresponding to September 5, 1994 at 0h00. As the original reference may be somewhat difficult to obtain we reproduce the initial conditions here. The masses, positions and velocities of the five bodies and physical units used can be found in Table \ref{table: solarsystem1} and Table \ref{table: solarsystem2}.

\begin{table}
\footnotesize
\begin{center}
\begin{tabular}{|c|c|}
\hline
Body & Mass [$M_\odot$]\\
\hline
Sun & 1.0\\
\hline
Jupiter & 9.54786104043$\times10^{-4}$\\
\hline
Saturn & 2.85583733151$\times10^{-4}$\\ 
\hline
Uranus & 4.37273164546$\times10^{-5}$\\ 
\hline
Neptune & 5.17759138449$\times10^{-5}$\\ 
\hline
\end{tabular}
\caption{The reference masses for bodies in the Solar system experiments. The masses of the inner planets can be taken into account by setting the Sun's mass into value of $1.00000597682\,M_\odot$.}
\label{table: solarsystem1}
\end{center}
\end{table}

\begin{table}
\begin{center}
\footnotesize
\begin{tabular}{|c|c|c|}
\hline
Body & Position [AU] & Velocity [AU/day]\\
\hline
\multirow{3}{4em}{} & 0.0 & 0.0\\ 
Sun & 0.0 & 0.0\\ 
& 0.0 & 0.0\\ 
\hline
\multirow{3}{3em}{} & -3.5023653 & +0.00565429\\ 
Jupiter & -3.8169847 & -0.00412490\\ 
& -1.5507963 & -0.00190589\\
\hline
\multirow{3}{4em}{} & +9.0755314 & +0.00168318\\ 
Saturn & -3.0458353 & +0.00483525\\ 
& -1.6483708 & +0.00192462\\
\hline
\multirow{3}{4em}{} & +8.3101420 & +0.00354178\\ 
Uranus & -16.2901086 & +0.00137102\\ 
& -7.2521278 & +0.00055029\\
\hline
\multirow{3}{4em}{} & +11.4707666 & +0.00288930\\ 
Neptune & -25.7294829  & +0.00114527\\ 
& -10.8169456 & +0.00039677\\
\hline
\end{tabular}
\caption{The reference positions and velocities for bodies in the Solar system experiments.}
\label{table: solarsystem2}
\end{center}
\end{table}

\subsection{Star cluster models}\label{appendix: plummer}

We use the N-body initial conditions code McLuster \cite{Kupper2011} for generating the star cluster models for this study. As input parameters we use the number of stars $N$ and the 3D half-mass radius $r_\mathrm{1/2}$ of the cluster model. The individual stellar masses are sampled from the initial mass function of \cite{Kroupa2001}. As the \frost{} code does not yet include stellar evolution we evolve the stellar population in time for $1$ Gyr using the SSE stellar evolutionary tracks of \citep{Hurley2000} so that the short-lived rapidly-evolving stars have collapsed into compact remnants. The maximum particle mass in the cluster models is thus $\sim 11\,M_\mathrm{\odot}$. In this study we include no primordial binary stars in the stellar population.

The stars are organised into a stellar cluster following the \cite{Plummer1911} model. The Plummer density-potential profile pair is defined as
\begin{equation}\label{eq: plummer-rho-phi}
\begin{split}
    \rho(r) &= \left( \frac{3 M}{4 \pi a^3} \right) \left(1+\frac{r^2}{a^2}\right)^{-5/2}\\
    \phi(r) &= -\frac{G M}{\sqrt{r^2+a^2}}
\end{split}
\end{equation}
Here $M$ is the total mass of the Plummer sphere and $a$ its scale radius. The cumulative mass profile $M(r)$ of the Plummer model is obtained from the density profile $\rho(r)$ the result being
\begin{equation}
    M(r) = M \left(1 + \frac{a^2}{r^2} \right)^{-3/2}.
\end{equation}
The stellar positions are generated using this cumulative mass profile. The relation between the half-mass radius $r_\mathrm{1/a}$ and the scale radius $a$ can also be computed from the cumulative mass profile. The result is
\begin{equation}
    r_\mathrm{1/2} = \frac{a}{\sqrt{2^{2/3}-1}} \approx 1.3048a.
\end{equation}
The stellar velocities are sampled using the distribution function $f(\mathcal{E})$ of the Plummer sphere which can be computed from the density-potential pair of Eq. \eqref{eq: plummer-rho-phi} using Eddington's method \citep{Binney2008}. Here $\mathcal{E} = -E = -1/2 v^2-\phi(r)$. The final formula for the Plummer distribution function can be written as
\begin{equation}
    f(\mathcal{E}) = 
    \begin{cases}
    \frac{24 \sqrt{2} a^2}{7 \pi^3 G^5 M^5} \mathcal{E}^{7/2} & \text{for } \mathcal{E}>0\\
    0 & \text{for } \mathcal{E} \leq 0.
    \end{cases}
\end{equation}


\bsp	
\label{lastpage}
\end{document}